\documentclass[twocolumn,aip,nofootinbib,jcp,reprint,amsmath,amssymb]{revtex4-2}

\usepackage{natbib,hyperref}
\usepackage{amsmath}
\usepackage{graphicx}
\usepackage{hyperref} 
\usepackage{dcolumn}
\usepackage{bm}
\usepackage{epsfig,color,xspace,multirow,xr,bbold}
\usepackage[all]{xy}
\usepackage{setspace}
\usepackage{url}
\usepackage{soul}
\usepackage[colorinlistoftodos]{todonotes}
\usepackage{longtable}
\usepackage{tabularx}
\usepackage{adjustbox}
\usepackage{threeparttable} 
\usepackage{natbib,hyperref}
\usepackage{float}
\usepackage{placeins}
\usepackage[utf8]{inputenc}
\usepackage[T1]{fontenc}

\definecolor{deepgreen}{RGB}{0,100,0}

\usepackage{xcolor}
\definecolor{deeppink}{RGB}{255,20,147}

\usepackage[normalem]{ulem}
\usepackage{xcolor}

\newcommand\notsotiny{\@setfontsize\notsotiny\@vipt\@viipt}
\usepackage{soul} 
\newcommand{\orcid}[1]{\href{https://orcid.org/#1}{\includegraphics[width=8pt]{orcid.png}}}
\begin{document}

\title{
Assessing excited-state geometry optimization strategies for adiabatic photophysical energies
}

\date{\today}

\author{
Amrita Bera, 
Atreyee Majumdar, 
Raghunathan Ramakrishnan$^*$ 
}
\email{ramakrishnan@tifrh.res.in} 
\thanks{\\ $^*$ Corresponding author}  
\affiliation{Tata Institute of Fundamental Research, Hyderabad 500046, India}


\begin{abstract}
\noindent
Accurate prediction of adiabatic $0$--$0$ excited-state energies is crucial for modeling molecular photophysical processes. 
Here, we benchmark computational strategies for evaluating
excited-state energies and singlet--triplet gaps obtained using different geometry-optimization strategies, including time-dependent density functional theory (TDDFT), spin-unrestricted Kohn--Sham (UKS) DFT for triplet states (${\rm T}_1$), and state-specific orbital-optimized UKS (ssUKS) DFT for singlet excited states (${\rm S}_1$). 
Zero-point vibrational energy corrections are evaluated consistently at the optimized geometries and combined with ADC(2) excitation energies for comparison with experimental anion photoelectron spectroscopy data for a representative set of molecules. 
Among the protocols considered, adiabatic $0$--$0$ energies evaluated at TDDFT-optimized ${\rm S}_1$ and ${\rm T}_1$ geometries show the best agreement with experiment, with a mean absolute error below 0.1~eV.
Replacing these geometries with UKS-optimized ${\rm T}_1$ and ssUKS-optimized ${\rm S}_1$ structures yields comparable accuracy. 
Vertical excitation energies are substantially more sensitive to the choice of geometry than the corresponding ${\rm S}_1$--${\rm T}_1$ gaps, which are comparatively more robust because of partial error cancellation.
As a larger case study, we examine rubrene and find that UKS/ssUKS-based geometries remain useful for evaluating singlet-fission energetics.
Overall, UKS/ssUKS-based workflows provide an efficient and accurate route to excited-state geometry optimization and to the evaluation of adiabatic 
$0$--$0$ energies for states with dominant single-determinant character.
\end{abstract}

\maketitle

\section{Introduction}\label{sec:introduction}
The photophysical performance of organic molecular light emitters and photovoltaic chromophores, including materials designed for thermally activated delayed fluorescence (TADF) and singlet fission (SF), is governed by excited-state energy differences~\cite{chen2018thermally,uoyama2012highly,smith2013recent}. The ${\rm S}_1 - {\rm T}_1$ gap controls the efficiency of reverse intersystem crossing (RISC) in TADF systems, while the condition $E({\rm S}_1) - 2E({\rm T}_1) \approx 0$  determines whether conventional SF is energetically allowed. 
Delayed fluorescence can also occur via the inverted singlet--triplet energy-gap mechanism (DFIST), characteristic of molecules with a negative 
${\rm S}_1-{\rm T}_1$ gap, often referred to as inverted singlet--triplet (INVEST) systems~\cite{de2019inverted,ehrmaier2019singlet,aizawa2022delayed,li2022organic,won2023inverted,majumdar2025unlocking,majumdar2025insights}. 
Higher-lying triplet states also play a role, since the condition 
$E({\rm T}_2) - 2E({\rm T}_1) \ge 0$ prevents the correlated triplet pair $^{1}({\rm T}_1 \cdots {\rm T}_1)$ from relaxing into ${\rm T}_2$, thereby suppressing triplet separation and promoting triplet--triplet annihilation (TTA) \cite{kondakov2015triplet}. In azulene, the ${\rm S}_2 - 2{\rm T}_1$ energy gap becomes relevant for assessing anti-Kasha SF pathways mediated by the second excited singlet state~\cite{veys2020computational,pino2024designing}. 
Collectively, these energetic relationships define the thermodynamic feasibility and directionality of a wide range of photophysical processes.

In vibronically resolved absorption or emission spectra, low-temperature phosphorescence measurements, and anion photoelectron spectroscopy (anion-PES), singlet and triplet excitation energies are most meaningfully defined as $0$--$0$ transition energies, corresponding to the energy difference between the vibrational ground states of two electronic states. By construction, these quantities incorporate both adiabatic structural relaxation and zero-point vibrational energy (ZPVE) corrections, and therefore reflect relaxed excited-state minima rather than purely vertical Franck--Condon transitions~\cite{fang2014method}. 
Since ZPVE contributions can amount to non-negligible contributions and may vary appreciably between the ground and excited states, their inclusion is often essential for chemically accurate comparisons with experiment, particularly when assessing small singlet--triplet gaps or near-degeneracy conditions relevant to TADF, INVEST, and SF systems~\cite{jacquemin2012basis,loos2019chemically}. 

Predicted adiabatic excitation energies, obtained by optimizing excited-state geometries and including ZPVE corrections, 
provide a closer theoretical counterpart to experimental observables~\cite{santoro2016going,ccaylak2021excited}. 
Such calculations often combine geometries and vibrational corrections obtained from density-functional theory (DFT) methods with more accurate electronic excitation energies evaluated using wavefunction-based approaches, particularly for larger systems where excited state geometry optimizations with correlated methods are impractical~\cite{bremond2018accuracy}. 
Nevertheless, the reliability of such composite strategies depends on the quality of the underlying excited-state geometries. While absorption spectra are often reasonably described using ground-state DFT geometries, emission energies and $0$--$0$ gaps are more sensitive to errors in excited-state geometries, which are typically larger for optimizations based on time-dependent density functional theory (TDDFT) than for ground-state DFT structures~\cite{jacquemin2018key,froitzheim2022either}.  
These observations highlight the critical roles of accurate excited-state geometries and adiabatic effects in quantitative photophysical modeling.

Electronic-structure approaches routinely employed in modeling molecular excited states include linear-response TDDFT (LR-TDDFT) and its Tamm--Dancoff approximation (LR-TDA). 
While LR-TDDFT is generally the method of choice for computing singlet excitation energies, the LR-TDA is often preferred in systems susceptible to triplet instabilities~\cite{peach2012overcoming,jindal2025comment}. 
Orbital-optimized excited-state DFT~\cite{hait2020excited,hait2021orbital} formalisms (also referred to as variational excited-state DFT methods~\cite{cheng2008rydberg,vigneshwaran2026variational}) based on non-Aufbau occupations, commonly formulated within a $\Delta$SCF framework~\cite{yang2024foundation,gilbert2008self,malis2025origin}, provide an alternative variational description of excited states and are most reliable when the targeted state is dominated by a single electronic configuration. 
The practical scope of such approaches has been further broadened by recent methodological developments, including implementations within the spin-restricted-open-shell Kohn--Sham (ROKS) framework together with improved maximum-overlap and orbital-tracking procedures~\cite{hait2016prediction,sinyavskiy2025bridging}. 

In addition, $\Delta$SCF-based excited-state methods have already been employed in application-driven studies ranging from excited-state \emph{ab initio} molecular dynamics to the prediction of ionization energies, electron affinities, and adiabatic triplet energetics in molecules relevant to organic light-emitting diodes (OLEDs)~\cite{vandaele2022deltascf,zheng2025theoretical}. 
However, because the excited-state wavefunction is still approximated by a single orbital-optimized determinant, this description can deteriorate for states with significant multiconfigurational character, where an explicit treatment of configuration mixing is essential.

For the lowest triplet excited state, ${\rm T}_1$, orbital optimization can often be carried out straightforwardly with spin-unrestricted Kohn--Sham (UKS) DFT. For the corresponding singlet excited state, however, the state-specific orbital-optimized UKS (ssUKS) approach relies on an open-shell determinant that is generally not a proper spin eigenfunction. The singlet case therefore provides a more stringent test of whether state-specific orbital-optimized DFT can yield reliable excited-state potential-energy surfaces and adiabatic energy differences. Given the distinct variational foundations of LR-TDA, UKS, and orbital-optimized excited-state DFT, these approaches may yield substantially different excited-state geometries and, consequently, different $0$--$0$ energy gaps, motivating a systematic assessment of computational strategies for adiabatic photophysical energy differences. 

From a practical perspective, orbital-optimized UKS and ssUKS excited-state calculations retain the SCF-type structure of ground-state Kohn--Sham theory, whereas LR-TDA geometry optimization requires repeated response calculations along the optimization pathway. In addition, LR-TDA optimizations often require computing and tracking several low-lying roots to maintain the correct state character throughout the optimization. Although the practical cost depends on the implementation and system size, these features make UKS/ssUKS-based geometry optimization especially attractive for larger molecules.

In this work, we benchmark multiple computational strategies for evaluating $0$--$0$ excitation energies and ${\rm S}_1$--${\rm T}_1$ gaps using a representative set of molecules: pentaazaphenalene (5AP), azulene, anthracene, fluoranthene, and pyruvic acid, for which precise ${\rm S}_1$ and ${\rm T}_1$ energies are available from cryogenic anion photoelectron spectroscopy and, where necessary, complementary high-resolution fluorescence measurements~\cite{wilson2024spectroscopic, vosskotter2015towards,baba2009structure,kregel2018photoelectron,chan1985spectroscopic,burrow2025singlet}. 
This benchmark set was selected to span chemically and photophysically distinct classes of current interest, including anti-Kasha behavior, singlet-fission-relevant chromophores, and carbonyl-containing systems, while avoiding open-shell radical-type species and diatomic molecules, for which the electronic-structure challenges and experimental interpretation can differ qualitatively. 
The chosen molecules, therefore, provide a compact but diverse test set for assessing single-determinant excited-state methods across different bonding motifs, relaxation patterns, and excited-state characters. 
By combining ground-state DFT, $\Delta$SCF-based excited-state optimizations, LR-TDA geometries, and unrestricted DFT triplet minima with ADC(2)-level excited-state modeling, we provide a systematic assessment of how methodological choices and geometry relaxation influence excited-state energy differences. 
Our focus is therefore not on comparing various formalisms for determining excitation energies, but on assessing how the choice of excited-state geometry optimization protocol propagates into ADC(2)-based adiabatic energies and singlet–triplet gaps. 
The validity of single-determinant excited-state formalisms is further examined through detailed molecular orbital (MO) analyses. 
To test whether the trends identified for the benchmark set remain useful in a larger, acene-derived system relevant to SF applications, we additionally examine rubrene.

\section{Computational Details}\label{sec:methods}
DFT calculations were carried out with ORCA~6.0.0~\cite{neese2018software,neese2025orca6} using the range-separated hybrid functional $\omega$B97X-D3 and the def2-TZVP basis set, together with the def2/J auxiliary basis within the RIJCOSX approximation.~\cite{vahtras1993integral,kendall1997impact}
Tight thresholds were used for both SCF convergence and geometry optimization, and all optimized structures were verified as stationary points by harmonic frequency analysis. With the same settings, excited-state geometry optimizations were performed with the LR-TDA formalism.
Single-point excited-state energies were evaluated with Q-Chem 6.0.2 \cite{krylov_gill_2013_qchem} using the ADC(2) method. Calculations were carried out with the cc-pVDZ, cc-pVTZ, aug-cc-pVDZ, and aug-cc-pVTZ basis sets, together with RI approximations where applicable. Unless otherwise stated, the results discussed in the main text correspond to the cc-pVDZ basis.

\begin{figure*}[!htbp]
    \centering
    \includegraphics[width=\textwidth]{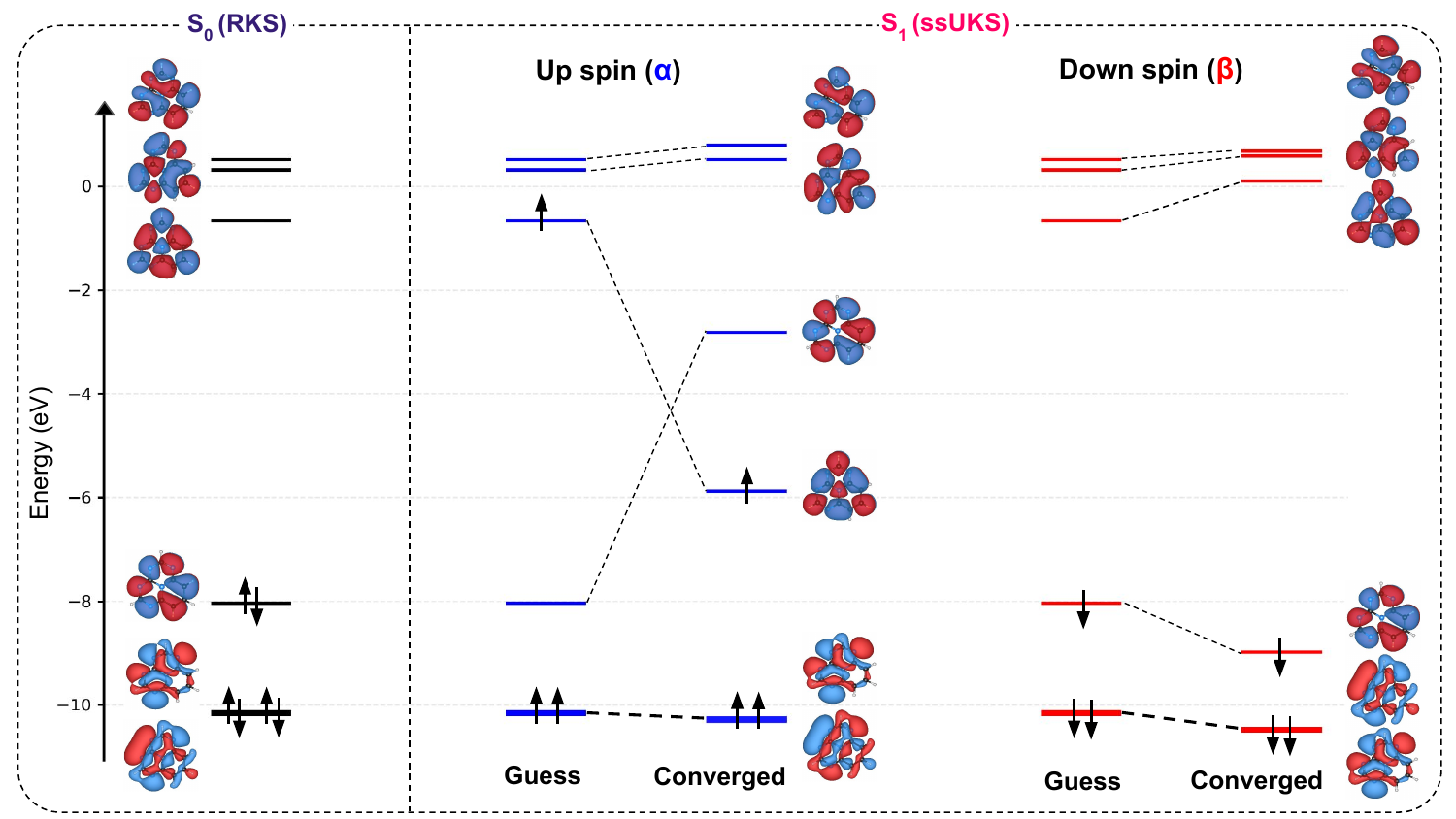}
    \caption{
    Frontier MOs of 5AP and their energies from orbital-optimized ground- and excited-state calculations. 
    MOs and their energies obtained for ${\rm S}_0$ are displayed, together with details from the ${\rm S}_1$ ssUKS calculation initialized from the ${\rm S}_0$ orbitals, shown separately for the $\alpha$ and $\beta$ spin channels.
    }
    \label{fig:5APssUKSMOs}
\end{figure*}

Ground-state (${\rm S}_0$) geometries were optimized with spin-restricted Kohn--Sham (RKS) DFT. The lowest singlet and triplet excited-state geometries, ${\rm S}_1$ and ${\rm T}_1$, were then optimized using two alternative protocols: (i) linear-response TDA--TDDFT (LR-TDA), and (ii) orbital-optimized unrestricted Kohn--Sham approaches. For ${\rm S}_1$, the latter corresponds to a state-specific UKS treatment with a fixed non-Aufbau occupation and is denoted here as state-specific UKS (ssUKS). 
Although the ${\rm S}_1$ ssUKS approach is implemented in ORCA using the \texttt{DeltaSCF} formalism, we avoid referring to it simply as $\Delta$SCF, since that term is often used more generally for excitation energies obtained as total-energy differences between independently optimized SCF solutions~\cite{ziegler1977calculation} rather than for excited-state calculation within the single-determinant KS-DFT approach.
For ${\rm T}_1$, a standard UKS calculation with triplet multiplicity was used. In all cases, harmonic vibrational frequencies were evaluated at the optimized geometries to obtain ZPVE corrections.

The lowest triplet state ${\rm T}_1$ is likewise treated using both LR-TDA and UKS calculations. In this case, the UKS description employs the standard Aufbau occupation appropriate for a triplet state. Accordingly, we do not use the term ssUKS for ${\rm T}_1$ and refer to this description simply as UKS, as there is no ambiguity in the electronic configuration and the calculation corresponds to a well-defined spin multiplicity of 3.

For pyruvic acid, an additional torsional scan was performed to examine the dependence of excited-state energies on the relevant dihedral coordinate. Starting from the optimized ${\rm S}_0$ geometry, all internal coordinates except the selected dihedral angle were frozen. 
The dihedral was varied from $0^\circ$ to $60^\circ$ in steps of  $10^\circ$. 
At each geometry, ${\rm S}_1$ and ${\rm T}_1$ excitation energies were determined with 
LR-TDA, ssUKS and UKS calculations.

\begin{figure*}[!htbp]
    \centering
    \includegraphics[width=\textwidth]{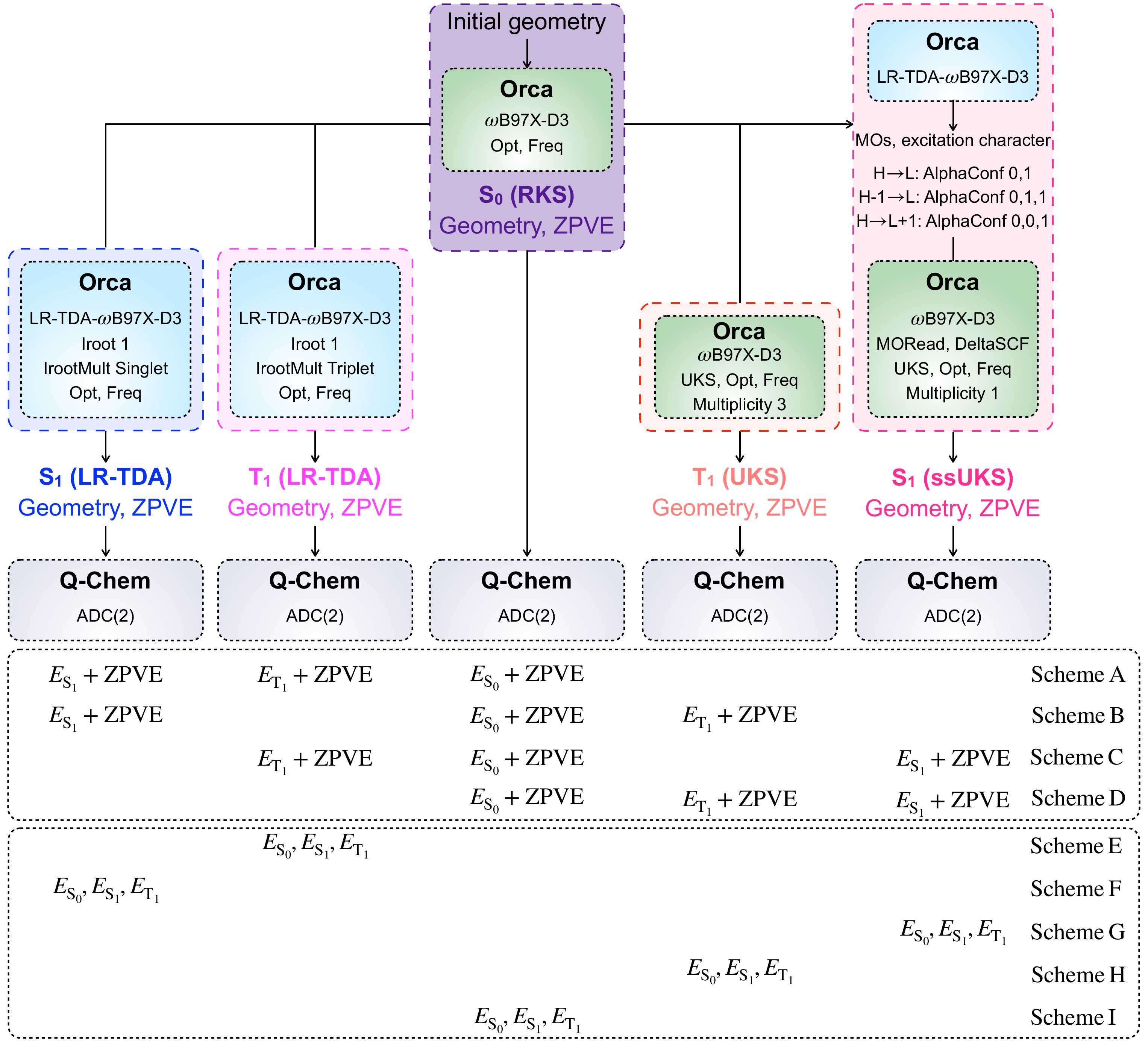}
    \caption{
Computational workflow used to evaluate adiabatic $0$--$0$ excited-state energies. The ground-state (${\rm S}_0$) geometry and the corresponding molecular orbitals (MOs) are used as initial guesses for all excited-state calculations. Singlet (${\rm S}_1$) and triplet (${\rm T}_1$) excited-state geometries are optimized using two approaches: (i) LR-TDA-TDDFT for both ${\rm S}_1$ and ${\rm T}_1$, and (ii) state-specific $\Delta$SCF/ssUKS for ${\rm S}_1$, and UKS for ${\rm T}_1$. In ssUKS ${\rm S}_1$ calculations, the electronic configuration of the $\alpha$-spin MOs is specified using the keyword {\tt AlphaConf 0,i,j,\ldots,k,1}, where {\tt 0} denotes the spin-MO from which an electron is excited, {\tt 1} denotes the spin-MO to which the electron is promoted, and {\tt i,j,k} represent intervening occupied or virtual spin-MOs, taking values of 1 for occupied and 0 for virtual orbitals. All optimized geometries are followed by harmonic frequency calculations to obtain ZPVE corrections for evaluating adiabatic $0$--$0$ energies. 
}
    \label{fig:workflow}
\end{figure*}

\subsection{State-specific UKS-DFT calculations}

Singlet excited-state (${\rm S}_1$) geometries were optimized using state-specific unrestricted Kohn--Sham density-functional theory (ssUKS-DFT), realized through the excited-state SCF/\texttt{DeltaSCF} machinery in ORCA. 
In this approach, a non-Aufbau determinant corresponding to the targeted excited-state configuration is optimized variationally, yielding MOs relaxed specifically for the chosen excited state. 
For a closed-shell reference, however, a singly excited singlet state is not represented by a single spin-unrestricted non-Aufbau determinant, but rather by the spin-adapted equal-weight combination of the $\alpha$- and $\beta$-excited configurations. Consequently, an ssUKS solution obtained by promoting an electron in only one spin channel corresponds to a broken-symmetry mixed-spin state and may exhibit different orbital-relaxation patterns in the two spin channels. This behavior is illustrated for 5AP in Fig.~\ref{fig:5APssUKSMOs}, which compares the frontier orbital energies of the ${\rm S}_0$ reference with those obtained from the ${\rm S}_1$ ssUKS calculation in the $\alpha$ and $\beta$ channels.
Analogous orbital-energy diagrams for azulene, anthracene, fluoranthene, and pyruvic acid are provided in Figs.~S1--S4  of the Supplementary Information (SI).

\begin{table*}[!htpb]
\centering
\caption{Experimental reference energies (in eV) compiled from cryogenic anion photoelectron spectroscopy (PES) and jet-cooled fluorescence excitation spectroscopy (FES). Values in parentheses denote the uncertainty in the last digit(s) of the stated value.}
\small
\begin{tabular}{l c c c c c l}
\hline
Molecule & ${\rm S}_1$ & ${\rm S}_2$ & ${\rm T}_1$ & ${\rm S}_1-{\rm T}_1$ & ${\rm S}_2-2{\rm T}_1$ & Source \\
\hline
5AP           &     $1.957(4)$  &          ---      &    $2.003(3)$  &           $-0.047(7)$  &          ---      &    Cryogenic anion PES~\cite{wilson2024spectroscopic}        \\
Azulene       &     $1.771$     &          $3.565$  &    $1.722$     &           $0.049$      &          $0.121$  &    Cryogenic anion PES~\cite{vosskotter2015towards}  \\
Anthracene    &     $3.433$     &          ---      &    $1.872(3)$  &           $1.561$      &          ---      &    ${\rm      S}_1$:                                    Jet-cooled  FES~\cite{baba2009structure};      ${\rm  T}_1$:  Cryogenic anion PES~\cite{kregel2018photoelectron}  \\
Fluoranthene  &     $3.126$     &          ---      &    $2.321(2)$  &           $0.805$      &          ---      &    ${\rm      S}_1$:                                    Jet-cooled  FES~\cite{chan1985spectroscopic};  ${\rm  T}_1$:  Cryogenic anion PES~\cite{kregel2018photoelectron}  \\
Pyruvic       acid  &           $3.25(3)$  &        ---  &           $2.962(8)$  &            $0.29(4)$  &        ---  &          Cryogenic                                 anion       PES~\cite{burrow2025singlet}       \\
Rubrene & $2.42(5)$ & --- & $1.16(5)$ & $1.26(7)$ & --- & Anion PES \cite{tsunoyama2017anion}\\
\hline
\end{tabular}
\label{tab:geom_benchmark}
\end{table*}

All ssUKS calculations were initialized from the converged ${\rm S}_0$ MOs of the corresponding ground-state calculation, which served as the reference for the subsequent LR-TDA analysis of excitation character. Since the ${\rm S}_0$  reference has a closed shell configuration, the $\alpha$ and $\beta$ orbitals are ideally identical, and the dominant excitation character may therefore be assigned from a common orbital manifold in both spin channels.
The desired excited-state occupation was imposed in the $\alpha$ channel using the \texttt{AlphaConf} keyword. 
For molecules whose lowest singlet excitation was identified from the LR-TDA calculation at the ${\rm S}_0$ geometry as predominantly ${\rm HOMO}\rightarrow{\rm LUMO}$ in character, the occupation pattern \texttt{AlphaConf 0,1} was used. When the relevant singlet excitation was instead dominated by a ${\rm HOMO}-1\rightarrow{\rm LUMO}$ transition, \texttt{AlphaConf 0,1,1} was employed. A ${\rm HOMO}\rightarrow{\rm LUMO}+1$ excitation was analogously represented by \texttt{AlphaConf 0,0,1}.
The ${\rm S}_1$ states of 5AP, azulene, anthracene, pyruvic acid, and rubrene were represented using the ${\rm HOMO}\rightarrow{\rm LUMO}$ occupation pattern, while fluoranthene was represented using the ${\rm HOMO}-1\rightarrow{\rm LUMO}$ pattern. In addition, the azulene ${\rm S}_2$ state was examined separately using both ${\rm HOMO}-1\rightarrow{\rm LUMO}$ and ${\rm HOMO}\rightarrow{\rm LUMO}+1$ occupation patterns.

During the SCF procedure, ORCA's maximum-overlap strategy was used to preserve the targeted excited-state occupation. To reduce root flipping and maintain consistent state tracking throughout the optimization, the initial reference determinant was retained using the default IMOM-based setup (\texttt{KEEPINITIALREF TRUE}). Orbital optimization was performed using the default second-order SCF procedure with an L-SR1 Hessian update, which is suitable for converging excited-state stationary points in orbital space.

\subsection{Workflow for excited-state calculations}

The overall computational protocol for evaluating vertical, adiabatic, and $0$--$0$ excitation energies is summarized in Fig.~\ref{fig:workflow}. Starting from the optimized ground-state ${\rm S}_0$ geometry and its MOs, excited-state geometries for ${\rm S}_1$ and ${\rm T}_1$ were obtained using either LR-TDA or orbital-optimized approaches, namely ssUKS for ${\rm S}_1$ and UKS for ${\rm T}_1$. Harmonic frequency calculations were then carried out at each optimized geometry to obtain ZPVE corrections, and single-point ADC(2) calculations were performed to evaluate the corresponding excitation energies.

The primary focus of the present work is on the $0$--$0$ schemes A and D, which compare two limiting protocols for excited-state geometry, while Schemes B and C are their mixed variants. 
In Schemes A--D, the adiabatic $0$--$0$ energies were calculated using the optimized excited-state geometries and the corresponding ZPVE corrections. 
On the other hand, Schemes E--I correspond to vertical excitation energies evaluated at the same set of optimized geometries. 
These vertical schemes were also examined because the required single-point energies are automatically obtained from the two main geometry protocols, namely Scheme-A and Scheme-D. 
The individual zero-point vibrational energy corrections used in the calculation of the $0$--$0$ energies are listed in Table~S4 of the SI.

ADC(2) calculations were performed at all optimized ${\rm S}_0$, ${\rm S}_1$, and ${\rm T}_1$ geometries in order to evaluate vertical, adiabatic, and $0$--$0$ excitation energies. For the ground state, the reference energy at the ${\rm S}_0$ minimum was taken directly from the MP2 component of the ADC(2) calculation at $R_{{\rm S}_0}$, and no separate MP2 calculation was performed. This value was used as the ground-state reference for evaluating adiabatic and $0$--$0$ excitation energies.

Adiabatic excitation energies were evaluated as
\begin{eqnarray}
E_{{\rm S}_1}^{\rm ad}
&=&
E_{{\rm S}_1}^{\rm ADC(2)}(R_{{\rm S}_1})
-
E_{{\rm S}_0}^{\rm MP2}(R_{{\rm S}_0}),
\\
E_{{\rm T}_1}^{\rm ad}
&=&
E_{{\rm T}_1}^{\rm ADC(2)}(R_{{\rm T}_1})
-
E_{{\rm S}_0}^{\rm MP2}(R_{{\rm S}_0}),
\end{eqnarray}
where $R_{{\rm S}_0}$, $R_{{\rm S}_1}$, and $R_{{\rm T}_1}$ denote the optimized geometries of the corresponding states.

The adiabatic $0$--$0$ energies were obtained by adding the corresponding zero-point vibrational energy corrections:
\begin{eqnarray}
E_{{\rm S}_1}^{0-0} &=&E_{{\rm S}_1}^{\rm ad} + \left[ {\rm ZPVE}_{{\rm S}_1} - {\rm ZPVE}_{{\rm S}_0} \right],
\\
E_{{\rm T}_1}^{0-0}&=&E_{{\rm T}_1}^{\rm ad} + \left[ {\rm ZPVE}_{{\rm T}_1} - {\rm ZPVE}_{{\rm S}_0} \right].
\end{eqnarray}

Here, the excited-state geometries $R_{{\rm S}_1}$ and $R_{{\rm T}_1}$ were taken either from LR-TDA optimizations or from ssUKS/UKS optimizations, depending on the protocol under consideration.

\begin{figure*}[!htpb]
    \centering
    \includegraphics[width=\linewidth]{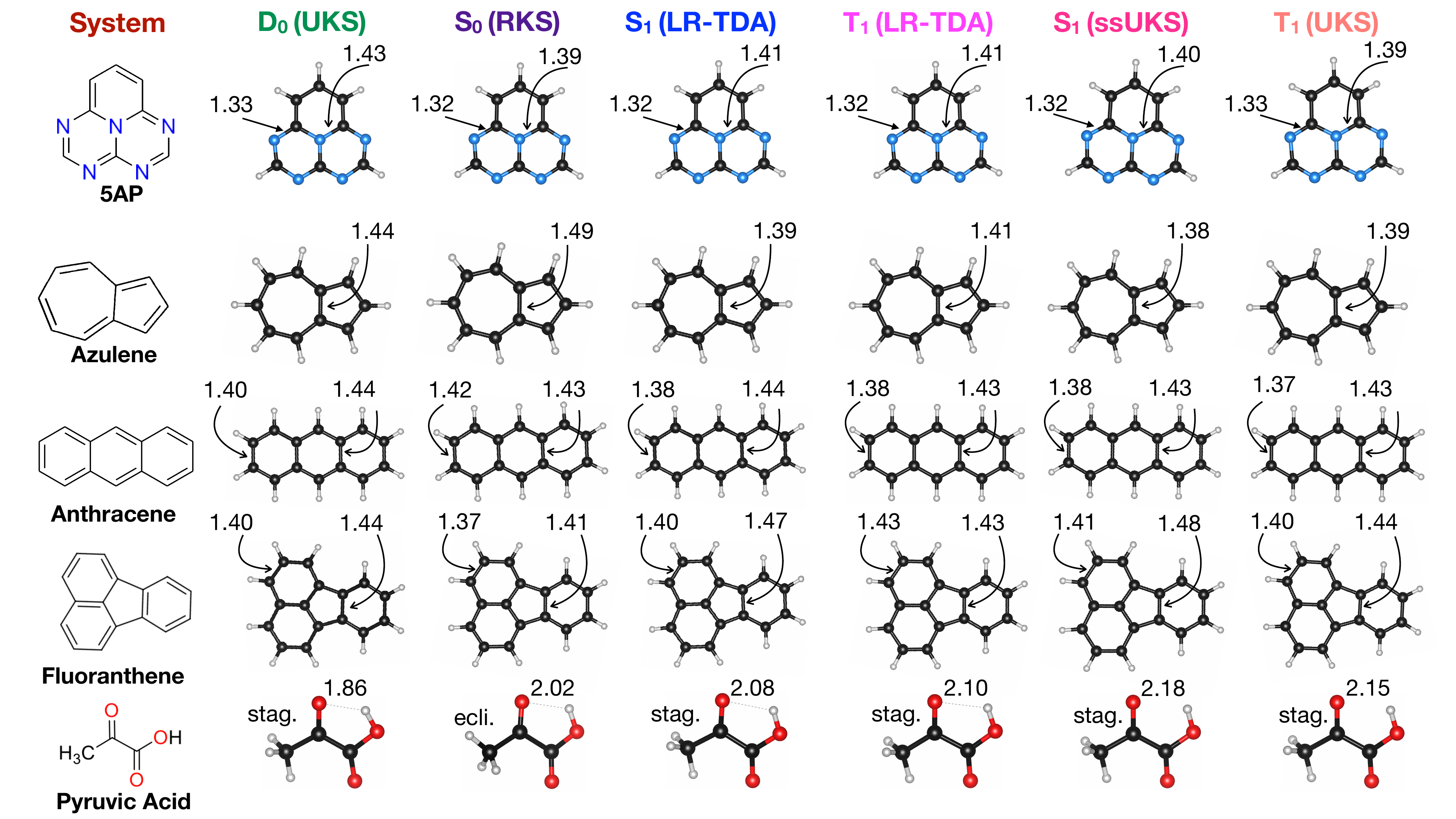}
    \caption{
    Chemical structures of 5AP, azulene, anthracene, fluoranthene, and pyruvic acid optimized at different electronic states and levels of theory: the anion ground state ${\rm D}_0$ (UKS, doublet); the neutral ground state ${\rm S}_0$ (RKS, singlet); the lowest singlet and triplet excited states ${\rm S}_1$ and ${\rm T}_1$ optimized using LR-TDA; the lowest singlet excited state ${\rm S}_1$ optimized using ssUKS with fixed non-Aufbau occupation; and the triplet state ${\rm T}_1$ optimized at the UKS level. Selected bond lengths are indicated.
    }
    \label{fig:molecules}
\end{figure*}

\section{Experimental Reference Data}\label{sec:dataset}
In this work, we consider a test set of five molecules with different photophysical characteristics (see Fig.~\ref{fig:molecules}).
The set comprises 
1,3,4,6,9b-pentaazaphenalene (5AP) as a representative inverted singlet--triplet (INVEST) system \cite{wilson2024spectroscopic,loos2023heptazine}, azulene as a prototypical anti-Kasha molecule \cite{dunlop2023excited,veys2020computational}, 
anthracene as a prototypical rigid polycyclic aromatic hydrocarbon relevant to singlet-fission, fluoranthene as a structurally related isomer with distinct excited-state relaxation behavior, 
and pyruvic acid as a flexible, non-rigid molecule of atmospheric relevance formed by oxidation of isoprene.
The experimental reference energies for all systems along with their sources are listed in Table~\ref{tab:geom_benchmark}.

For all systems, photoelectron spectroscopy probes neutral electronic states via photodetachment from the anion ground state ${\rm D}_0^{\rm anion}(\nu=0)$, providing access to neutral states that carry significant one-electron detachment character. This enables direct spectroscopic determination of the neutral ground state through the ${\rm S}_0^{\rm neutral}(\nu=0) \leftarrow {\rm D}_0^{\rm anion}(\nu=0)$ transition, and, depending on the molecule, selected low-lying singlet and triplet excited states of the neutral molecule on an equal footing.

For 5AP, high-resolution cryogenic anion photoelectron spectroscopy of the radical anion yielded the ${\rm S}_0^{\rm neutral}(\nu=0) \leftarrow {\rm D}_0^{\rm anion}(\nu=0)$ transition at an electron binding energy of $1.235(2)$~eV, corresponding to the adiabatic electron affinity~\cite{wilson2024spectroscopic}. Two closely spaced features at $3.192(4)$~eV and $3.239(3)$~eV were assigned to the $0$--$0$ origin transitions of the ${\rm S}_1$ and ${\rm T}_1$ states, yielding adiabatic term energies of $1.957(4)$~eV and $2.003(3)$~eV relative to ${\rm S}_0$, and establishing an inverted singlet--triplet gap of $\Delta E_{\rm ST} = -0.047(7)$~eV. The ${\rm S}_1$ assignment was independently confirmed by electronic absorption spectroscopy in an annealed 20~K argon matrix.

In the case of azulene, photodetachment photoelectron spectroscopy of the cryogenically cooled radical anion yielded an electron affinity of $0.790(8)$~eV for the ${\rm S}_0^{\rm neutral} \leftarrow {\rm D}_0^{\rm anion}$ transition~\cite{vosskotter2015towards,schiedt2000microsolvation}. Peaks at $1.722$~eV and $1.771$~eV were assigned to the ${\rm T}_1$ and ${\rm S}_1$ states, corresponding to an unusually small ${\rm S}_1$--${\rm T}_1$ splitting of $0.049$~eV ($\sim 395~\mathrm{cm^{-1}}$). Higher excited states, including ${\rm S}_2$ at $3.565$~eV, were also observed, with assignments supported by DFT/MRCI calculations and Franck--Condon simulations \cite{vosskotter2015towards}.

The triplet-state energies of anthracene and fluoranthene were determined using slow electron velocity-map imaging photoelectron spectroscopy of their cryogenically cooled radical anions~\cite{kregel2018photoelectron}. For anthracene, the ${\rm S}_0^{\rm neutral} \leftarrow {\rm D}_0^{\rm anion}$ $0$--$0$ transition was observed at $4290(24)\,\mathrm{cm^{-1}}$, while the ${\rm T}_1^{\rm neutral} \leftarrow {\rm D}_0^{\rm anion}$ transition occurred at $19\,387(25)\,\mathrm{cm^{-1}}$, yielding an adiabatic ${\rm S}_0$--${\rm T}_1$ splitting of $15\,097(25)\,\mathrm{cm^{-1}}$, i.e., 1.872(3)~eV. For fluoranthene, the corresponding ${\rm S}_0$ and ${\rm T}_1$ $0$--$0$ transitions were measured at $6108(15)\,\mathrm{cm^{-1}}$ and $24\,829(15)\,\mathrm{cm^{-1}}$, respectively, giving an adiabatic splitting of $18\,721(15)\,\mathrm{cm^{-1}}$, i.e., 2.321(2)~eV.

As the ${\rm S}_1$ states of anthracene and fluoranthene lie substantially higher in energy than ${\rm T}_1$ and carry negligible one-electron detachment character, they are not observed in the photoelectron spectra~\cite{kregel2018photoelectron}. 
However, we emphasize that the accessibility of excited electronic states in photoelectron spectroscopy is governed by their one-electron detachment character, as quantified by the Dyson orbital overlap between the anion ground state and the neutral final state~\cite{melania2007dyson}, rather than by their absolute excitation energy. 
In anthracene and fluoranthene, the ${\rm S}_1$ state is a genuine $\pi\!\rightarrow\!\pi^\ast$ excitation with negligible Dyson overlap and is therefore inaccessible in PES. 
Their ${\rm S}_0$--${\rm S}_1$ excitation energies were therefore taken from fluorescence excitation measurements, yielding $27\,687.15~\mathrm{cm^{-1}}$ (3.433~eV) for anthracene~\cite{baba2009structure} and $25\,216.9~\mathrm{cm^{-1}}$ (3.126~eV) for fluoranthene~\cite{chan1985spectroscopic}.

For pyruvic acid, anion photoelectron spectra recorded at multiple photon energies enabled measurement of adiabatic detachment energies to the ${\rm S}_0$, ${\rm T}_1$, and ${\rm S}_1$ states \cite{burrow2025singlet}. The ${\rm S}_0^{\rm neutral} \leftarrow {\rm D}_0^{\rm anion}$ transition was measured at $0.901(5)$~eV, while detachment to ${\rm T}_1$ and ${\rm S}_1$ occurred at $3.863(5)$~eV and $4.15(3)$~eV, respectively \cite{burrow2025singlet}. These values correspond to adiabatic term energies of $E({\rm T}_1)=2.962(8)$~eV and $E({\rm S}_1)=3.25(3)$~eV relative to ${\rm S}_0$, yielding an ${\rm S}_1$--${\rm T}_1$ gap of $0.29(4)$~eV. Franck--Condon simulations based on DFT and TDA--TDDFT calculations support the spectral assignments and confirm the adiabatic character of the extracted gaps.

\section{Results and Discussion}\label{sec:results}
\subsection{Choice of basis set}

A summary of the basis-set dependence of the ADC(2) adiabatic $0$--$0$ energies is provided in Tables~S1--S2 of
the Supporting Information (SI). 
In this comparison, the ${\rm S}_0$ geometry was optimized at the RKS level, while the ${\rm S}_1$ and ${\rm T}_1$ geometries entering the reported $0$--$0$ energies were obtained from LR-TDA--TDDFT optimizations. Subsequently, single-point excitation energies were determined with ADC(2) level using cc-pVDZ, cc-pVTZ, aug-cc-pVDZ, and aug-cc-pVTZ basis sets. 
The absolute ${\rm S}_1$ and ${\rm T}_1$ excitation energies show only modest basis-set dependence, typically on the order of $\sim 0.1$~eV, with diffuse augmentation having the largest effect for the more delocalized $\pi$-systems. In contrast, the ${\rm S}_1$--${\rm T}_1$ gaps are considerably less sensitive to basis-set choice, reflecting substantial cancellation of basis-set errors between the singlet and triplet states. Since the qualitative energetic trends are unchanged across the basis sets examined, ADC(2)/cc-pVDZ was selected for the subsequent analysis as a computationally economical level that still provides a reliable description of the singlet--triplet gaps. 

\subsection{Structural and electronic characterization of the benchmark set}
Fig.~\ref{fig:molecules} shows the benchmark molecules together with the optimized structures obtained for ${\rm D}_0$, ${\rm S}_0$, ${\rm S}_1$, and ${\rm T}_1$ using the different electronic-structure approaches considered in this work. For the rigid $\pi$-conjugated systems 5AP, azulene, anthracene, and fluoranthene, the structural differences between LR-TDA and ssUKS/UKS minima are generally modest, although clear state-dependent bond-length redistribution is observed, particularly for anthracene and fluoranthene. In contrast, pyruvic acid shows much larger geometry changes across electronic states and methods, reflecting its greater conformational flexibility and stronger relaxation effects. Thus, the benchmark set spans both electronically and structurally distinct classes of molecules and provides a suitable basis for assessing the feasibility and accuracy of single-reference excited-state approximations for geometry optimization. Since ssUKS and UKS optimizations are computationally comparable in cost to ground-state optimizations, they are especially attractive for larger molecules. The goal here is therefore to examine how well such approaches reproduce excited-state geometries relative to LR-TDA, and how these differences affect the resulting adiabatic $0$--$0$ energies and singlet--triplet gaps.

\begin{figure}[!htbp]
    \centering
    \includegraphics[width=\columnwidth]{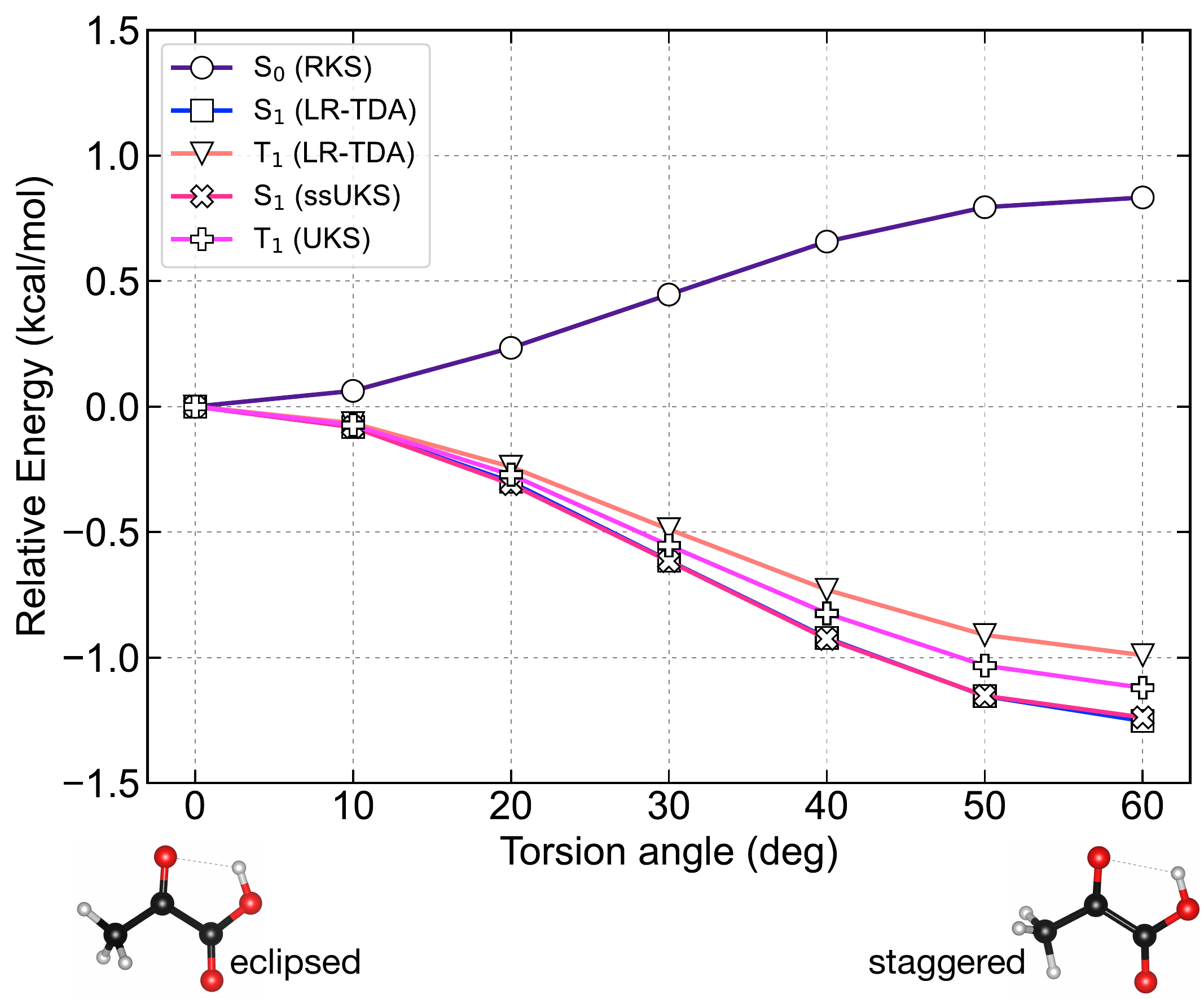}
    \caption{
Torsional energy profiles of pyruvic acid along the ${\rm CH_3}$ torsional coordinate for the ground state ${\rm S}_0$ and the lowest excited singlet and triplet states. The ${\rm S}_1$ and ${\rm T}_1$ profiles are shown as obtained from LR-TDA, ssUKS, and UKS calculations.
}
    \label{fig:torsionpyruvicacid}
\end{figure}

\begin{figure*}[!htbp]
    \centering
    \includegraphics[width=1.\textwidth]{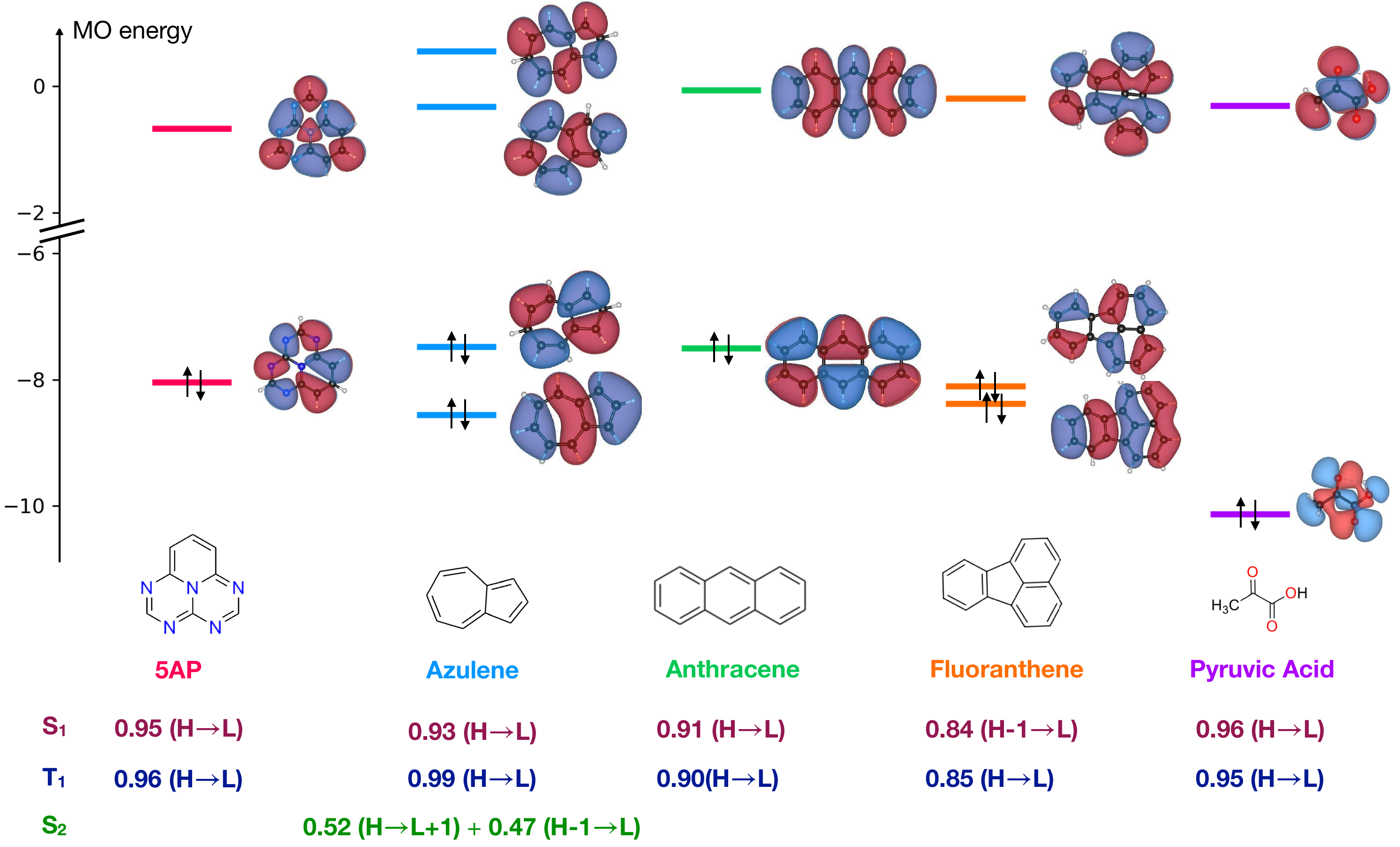}
\caption{Frontier molecular orbitals and dominant orbital contributions to the low-lying excited states of the benchmark systems. Shown are the relevant occupied and virtual orbitals together with their orbital energies from the ground-state calculation. For each molecule, the leading configurations and their weights are given for the ${\rm S}_1$ and ${\rm T}_1$ states, and for azulene also for ${\rm S}_2$.
}
    \label{fig:MOsbenchmarkset}
\end{figure*}

Among the benchmark systems, pyruvic acid is particularly useful for assessing whether single-determinant excited-state methods can reproduce the behavior of flexible excited-state potential-energy surfaces at a level comparable to LR-TDA, because its low-energy torsional coordinate introduces appreciable conformational sensitivity. This is illustrated by the torsional profiles in Fig.~\ref{fig:torsionpyruvicacid}. 
The numerical values underlying these torsional profiles are collected in Table~S3 of the SI. 

Along the relevant dihedral coordinate, the ground state ${\rm S}_0$ favors the eclipsed arrangement, whereas the low-lying excited states show the opposite trend and are stabilized toward the staggered configuration. Importantly, the ${\rm S}_1$ profiles obtained with LR-TDA and ssUKS are essentially coincident over the full torsional range, indicating that both methods describe the singlet excited-state conformational dependence in a nearly identical manner. For ${\rm T}_1$, the UKS and LR-TDA curves are likewise very similar, with UKS yielding only a slightly greater stabilization at larger torsional angles, by about 0.2 kcal/mol. Thus, despite the soft torsional coordinate and the stronger structural relaxation in pyruvic acid, the orbital-optimized approaches reproduce the same qualitative excited-state preference as LR-TDA and differ only minimally. Overall, Fig.~\ref{fig:torsionpyruvicacid} highlights why flexible molecules provide a more stringent test of excited-state geometry methods than rigid aromatic systems, since even small differences in the torsional profile can propagate into the adiabatic and $0$--$0$ excitation energies.

Spin contamination remains a possible source of systematic error, since it can influence not only total energies but also optimized geometries and harmonic vibrational frequencies, with the latter depending sensitively on the geometrical variation of $\langle \hat{S}^2\rangle$ rather than on its absolute value alone.\cite{jensen1990remarkable} For this reason, the variation of $\langle \hat{S}^2\rangle$ along the pyruvic acid torsional coordinate was also examined. Table~S3 shows that $\langle \hat{S}^2\rangle$ remains nearly constant at $\approx 1.0$ along the torsional scan, indicating that the close agreement between the LR-TDA and ssUKS torsional profiles is not associated with any significant torsion-dependent change in spin contamination. More generally, the absence of large geometry-dependent variations in $\langle \hat{S}^2\rangle$ suggests that the ssUKS description remains sufficiently stable along the relevant coordinates.

\begin{table*}[!htpb]
\centering
\caption{
Adiabatic $0$--$0$ excitation energies (in eV) for azulene, 5AP, pyruvic acid, anthracene, and fluoranthene computed at the ADC(2)/cc-pVDZ level using two geometry protocols. In Scheme-A, the ${\rm S}_1$ and ${\rm T}_1$ geometries and ZPVE corrections were obtained from LR-TDA optimizations, whereas in Scheme-D the ${\rm S}_1$ and ${\rm T}_1$ geometries and ZPVE corrections were obtained from ssUKS and UKS optimizations, respectively. 
In both schemes, the ground-state ${\rm S}_0$ geometry and the corresponding zero-point vibrational energy were obtained at the RKS--$\omega$B97X-D3/def2-TZVP level. Also given are the 
Experimental reference values for comparison. 
}
\label{tab:00vsUKS}
\renewcommand{\arraystretch}{1.2}
\begin{tabular}{l cccc cccc   ccc}
\hline
\multicolumn{1}{l}{System} & 
\multicolumn{3}{l}{Scheme-A} & &
\multicolumn{3}{l}{Scheme-D} & &
\multicolumn{3}{l}{Exp.}\\
\cline{2-4} \cline{6-8} \cline{10-12}
&  
\multicolumn{1}{l}{S$_1$} & \multicolumn{1}{l}{T$_1$} & \multicolumn{1}{l}{S$_1$-T$_1$} & &  
\multicolumn{1}{l}{S$_1$} & \multicolumn{1}{l}{T$_1$} & \multicolumn{1}{l}{S$_1$-T$_1$} & &  
\multicolumn{1}{l}{S$_1$} & \multicolumn{1}{l}{T$_1$} & \multicolumn{1}{l}{S$_1$-T$_1$}\\
\hline
5AP           &$1.94$ &$2.08$ & $-0.14$& & $1.88$ & $2.00$ & $-0.12$ & & $1.957$ & $2.003$ & $-0.047$\\
Azulene       &$1.84$ & $1.80$ & $0.04$ & & $1.79$ & $1.78$ & $0.01$ & & $1.771$  & $1.722$ & $0.049$ \\
Anthracene    &$3.39$ & $1.86$ & $1.53$ & & $3.36$ & $1.86$ & $1.50$ & & $3.433$ &  $1.872$ & $1.561$ \\
Fluoranthene  &$3.20$ & $2.36$ & $0.84$ & &  $3.16$ & $2.39$ & $0.76$ & & $3.126$ & $2.321$ & $0.805$ \\
Pyruvic acid  &$3.15$ & $2.75$ & $0.40$ & & $3.11$ & $2.71$ & $0.40$ & &  $3.25$ & $2.962$ & $0.29$\\
\hline
\end{tabular}
\end{table*}

For 5AP, azulene, anthracene, and pyruvic acid, both the ${\rm S}_1$ and ${\rm T}_1$ states are dominated by a single frontier-orbital promotion, with weights of about 0.9 or greater for the leading configuration (Fig.~\ref{fig:MOsbenchmarkset}). This dominant single-excitation character makes these states well-suited to ssUKS and UKS treatments, because the corresponding excited-state relaxation can be described largely by a single non-Aufbau occupation pattern.
Likewise, Fluoranthene is largely single-configurational, but its lowest singlet and triplet excitations are dominated by a ${\rm HOMO}-1 \rightarrow {\rm LUMO}$ promotion rather than the nominal ${\rm HOMO}\rightarrow{\rm LUMO}$ excitation. This highlights the need for careful specification of the targeted occupation pattern, as discussed in \ref{sec:methods}.

\subsection{Assessment of geometry protocols for adiabatic and vertical excitation energies}

Table~\ref{tab:00vsUKS} compares ADC(2)/cc-pVDZ adiabatic $0$--$0$ energies obtained with two excited-state geometry protocols: Scheme-A, based on LR-TDA geometries for both ${\rm S}_1$ and ${\rm T}_1$, and Scheme-D, based on ssUKS and UKS geometries for ${\rm S}_1$ and ${\rm T}_1$, respectively, with the same RKS ${\rm S}_0$ reference used in both cases. 
The corresponding ZPVE corrections are collected in Table~S4 of the SI.
Both schemes yield similar results across the benchmark set, indicating that replacing LR-TDA excited-state geometries with orbital-optimized single-reference geometries does not qualitatively alter the predicted adiabatic energetics. 
${\rm S}_1-{\rm T}_1$ gaps from both Scheme-A and Scheme-D are in good agreement with the experimental values, and capture the INVEST nature of 5AP.
These results indicate that ssUKS/UKS geometries provide a useful low-cost alternative to LR-TDA minima for evaluating adiabatic excitation energies in states of predominantly single-determinant character.
The close similarity between Scheme-A and Scheme-D is further supported by the basis-set comparison in Figs.~S5--S7 (see SI), which shows that both protocols display the same overall convergence behavior from cc-pVDZ to cc-pVTZ.

\begin{table*}[!htpb]
\centering
\caption{
Statistical analysis of errors in $0$–$0$, adiabatic, and vertical excitation energies (in eV) computed with ADC(2)/cc-pVDZ using different combinations of ${\rm S}_0$, ${\rm S}_1$, and ${\rm T}_1$ geometry optimization strategies. Ground-state geometries were obtained using RKS–DFT, and excited-state geometries optimized using LR-TDA–TDDFT, ssUKS, or UKS, as indicated. Errors are reported with respect to experimental reference values for azulene, 5AP, pyruvic acid, anthracene, and fluoranthene. 
MAE and SDE denote the mean absolute error and standard deviation of the error. Values $<0.1$ eV are in bold face.
}
\label{tab:geomeffect}
\begin{tabular}{c ll lll llll }
\hline
\multicolumn{1}{l}{Scheme} & 
\multicolumn{1}{l}{Excitation type} & 
\multicolumn{3}{l}{Geometry optimization method} & &
\multicolumn{4}{l}{MAE (SDE)} \\
\cline{3-5} 
\cline{7-10} 
& 
&  \multicolumn{1}{l}{S$_0$} & \multicolumn{1}{l}{S$_1$} & \multicolumn{1}{l}{T$_1$} &
&  \multicolumn{1}{l}{S$_1$} & \multicolumn{1}{l}{T$_1$} & \multicolumn{1}{l}{S$_1$-T$_1$} & \multicolumn{1}{l}{combined} \\
\hline
A & 0-0 & RKS & LR-TDA & LR-TDA & & ${\bf0.06\,(0.07)}$ & ${\bf0.08}$ ($0.11$) & ${\bf0.06\,(0.07)}$ & ${\bf0.07(0.08)}$$^a$   \\
B & 0-0 &  RKS & LR-TDA &UKS & &  ${\bf0.06\,(0.07)}$ & ${\bf0.08}$ ($0.12$) & ${\bf0.04\,(0.06)}$  & ${\bf0.06\,(0.09)}$\\
C & 0-0 & RKS& ssUKS & LR-TDA &  &  ${\bf 0.07\,(0.07)}$ & ${\bf0.08}$ ($0.11$) & ${\bf 0.07\,(0.08)}$ & ${\bf 0.08\,(0.09)}$  \\
D & 0-0 & RKS&  ssUKS&  UKS  & & $\bf {0.07\,(0.07)}$ & ${\bf0.08}$\,($0.12$) & ${\bf 0.07\,(0.07)}$   & ${\bf 0.07\,(0.09)}$$^b$   \\
E & Vertical &---&---&  LR-TDA &  & 0.14 (0.16) & {0.10} (0.13) & 0.14 (0.16)  & 0.13 (0.15) \\
F & Vertical &---& LR-TDA &---& &  0.13 ($\bf0.07$) & 0.12 (0.14) & 0.14 (0.12)  &  0.13 (0.12)\\
G & Vertical &---&---& UKS &  & 0.24 (0.15) & 0.24 ($\bf0.09$) & {$0.10$} (0.11) & 0.19 (0.17) \\
H & Vertical&---& ssUKS &---&  & 0.25 ${\bf (0.09)}$ & 0.17 (0.17) & 0.13 (0.11) & 0.18 (0.15) \\
I & Vertical & RKS &--- &---& & 0.48 (0.21) & 0.49 (0.19) & 0.13 (0.13)  & 0.37 (0.30) \\
\hline
\end{tabular}
\begin{tablenotes}
 \item  \footnotesize{$^a$ Without ZPVE corrections, combined MAE(SDE) value of adiabatic energies is $0.10$ (0.11) eV.}
  \item  \footnotesize{$^b$ Without ZPVE corrections, combined MAE(SDE) value of adiabatic energies is $0.10$ (0.11) eV.}
 \end{tablenotes}
\end{table*} 

Table~\ref{tab:geomeffect} summarizes the mean absolute errors (MAE) and standard deviation of the errors (SDE) 
of various schemes (A--I, see Fig.~\ref{fig:workflow}). The best overall performance is obtained for the $0$--$0$ schemes (A--D) based on LR-TDA ${\rm S}_1$ geometries, with Scheme-A (LR-TDA/LR-TDA) and Scheme-B (LR-TDA/UKS) yielding combined MAEs of 0.07 and 0.06 eV, respectively. This indicates that UKS provides a practically equivalent description of the ${\rm T}_1$ minimum, whereas replacing LR-TDA by ssUKS for ${\rm S}_1$ (Schemes C and D) leads to only a modest loss of accuracy. 

In contrast, all vertical schemes (E--I) are noticeably worse, and vertical excitation energies evaluated at the ${\rm S}_0$ geometry (Scheme-I) are by far the least reliable, with MAEs approaching 0.5 eV for both ${\rm S}_1$ and ${\rm T}_1$. The ${\rm S}_1$--${\rm T}_1$ gaps are systematically less sensitive than the absolute excitation energies, reflecting favorable cancellation of errors between the two excited states. 

Finally, the nearly unchanged error statistics upon omission of ZPVE corrections for Schemes A and D (see Table~\ref{tab:geomeffect} footnote) show that the main conclusions are governed by the excited-state geometries, while inclusion of ZPVE is still required for direct comparison to experimental $0$--$0$ energies.
Accordingly, Schemes A and D define the most informative limiting comparison between fully LR-TDA and fully orbital-optimized excited-state geometry optimization protocols, while Schemes B and C show the expected intermediate behavior.

It is important to emphasize that the present assessment does not rely on direct excited-state energies from ssUKS or UKS. Rather, these methods are used to generate excited-state geometries and ZPVE corrections, while the final excitation energies are evaluated at the ADC(2) level. Accordingly, any spin contamination in the orbital-optimized DFT states is expected to affect the results primarily through the quality of the underlying excited-state potential-energy surfaces and the associated ZPVE corrections. The expectation values of the total spin operator, $\langle \hat{S}^2\rangle$, for the optimized ssUKS ${\rm S}_1$ solutions are listed for all five benchmark systems in Table~S5 of the SI. As expected for a broken-symmetry open-shell description of an excited singlet state, the ssUKS solutions are not pure spin eigenfunctions and show $\langle \hat{S}^2\rangle$ values between 1.012 and 1.172. The largest deviations occur for 5AP and azulene, whereas anthracene, fluoranthene, and pyruvic acid remain somewhat closer to the nominal broken-symmetry reference value. Despite this, the ADC(2)-based adiabatic $0$--$0$ energies obtained with Scheme-D remain very similar to those from the fully LR-TDA-based Scheme-A, indicating that the spin contamination in the ssUKS reference is not large enough to significantly alter the relevant excited-state geometries or the resulting adiabatic energetics for the present benchmark set.

Azulene highlights an important limitation of the present approach for excited singlet states. 
Its characteristic anti-Kasha behavior is associated with emission from the higher-lying ${\rm S}_2$ state rather than from ${\rm S}_1$.~\cite{beer1955anomalous,dunlop2023excited,bearpark1996azulene,pino2024designing} 
Azulene ${\rm S}_2$ is not dominated by a single orbital promotion, but instead shows substantial mixing of ${\rm HOMO}\rightarrow{\rm LUMO}+1$ and ${\rm HOMO}-1\rightarrow{\rm LUMO}$ configurations (see Fig.~\ref{fig:MOsbenchmarkset}). 
This mixed character lies beyond the scope of the ssUKS description as illustrated by additional calculations summarized in Table~S6.

At the LR-TDA-based Scheme-A level, the predicted ${\rm S}_2$, ${\rm T}_1$, and ${\rm S}_2-2{\rm T}_1$ values are 3.719, 1.802, and 0.115 eV, respectively, in very good agreement with the experimental ${\rm S}_2-2{\rm T}_1$ value of 0.121 eV (see Table~\ref{tab:geom_benchmark}). In Scheme-D, the ${\rm T}_1$ energy remains nearly unchanged at 1.777 eV, while the predicted ${\rm S}_2$ energy are predicted as 3.870 and 3.873 eV for the ${\rm HOMO}\rightarrow{\rm LUMO}+1$ and ${\rm HOMO}-1\rightarrow{\rm LUMO}$ non-Aufbau occupation patterns, respectively. 
The corresponding ${\rm S}_2-2{\rm T}_1$ values, 0.316 and 0.319 eV, are therefore also very similar to one another. Although both 
variants of Scheme-D are less accurate than Scheme-A and experiment, they are still closer to the adiabatic reference than the vertical estimates based on ${\rm S}_0$ or ${\rm T}_0$  geometries, which give ${\rm S}_2-2{\rm T}_1=0.631$ eV for Scheme-G and $-0.634$ eV for Scheme-I. All together, these results indicate that the present ssUKS-based protocol does not provide a robust quantitative description of the ${\rm S}_2$ state of azulene with substantial configuration mixing, yet, the relaxed-state energetics remain qualitatively more reasonable than purely vertical estimates.

\subsection{Rubrene as a larger molecule case study}

Rubrene provides a useful larger-molecule test because full LR-TDA excited-state geometry optimizations become substantially more demanding for systems of this size. Accordingly, only Scheme-D and the corresponding vertical schemes were examined, allowing us to test whether the performance of the ssUKS/UKS-based protocol identified for the smaller benchmark set remains valid for a more extended $\pi$-conjugated chromophore.
Rubrene is a known singlet-fission candidate~\cite{ma2012singlet,sutton2017singlet} for which the energy condition $E({\rm S}_1)-2E({\rm T}_1)\ge 0$ is experimentally delicate. Gas-phase anion photoelectron spectroscopy reported $E({\rm S}_1)=2.42\pm0.05$ eV and $E({\rm T}_1)=1.16\pm0.05$ eV for the isolated molecule, placing rubrene close to the singlet-fission threshold with $E({\rm S}_1)-2E({\rm T}_1)=0.1$~eV. In the same study, the vibronic profiles of the neutral ${\rm S}_0$, ${\rm T}_1$, and ${\rm S}_1$ states were found to be similar, indicating relatively small structural displacement upon photoexcitation, and the twisted $D_2$ form was identified as the more stable molecular structure.\cite{tsunoyama2017anion}
Against this background, the excitation energies collected in Table~\ref{tab:rubrene} provide a direct test of whether the conclusions drawn from the smaller benchmark set remain valid for a larger singlet-fission chromophore.

\begin{figure*}[!htbp]
    \centering
    \includegraphics[width=0.7\textwidth]{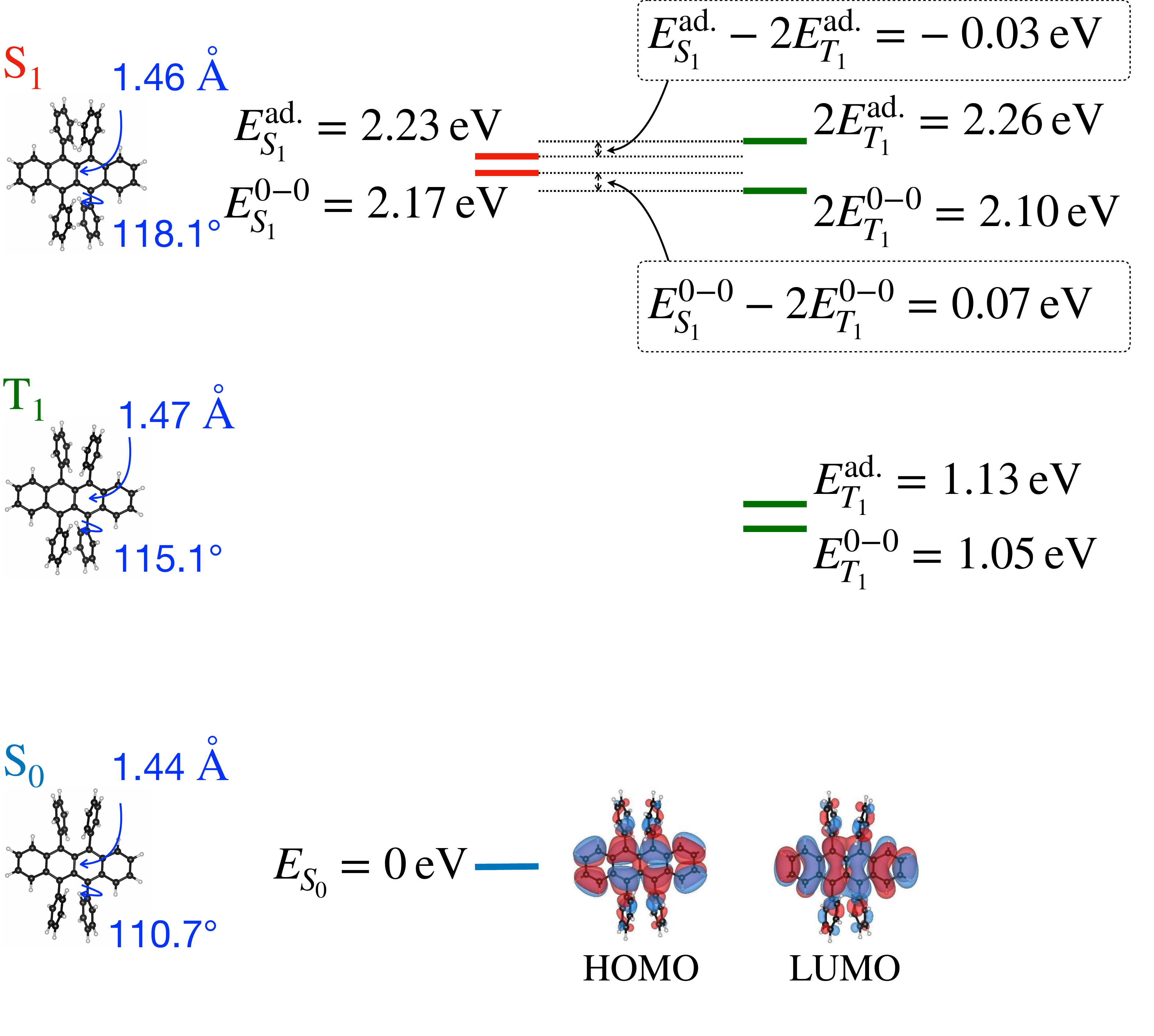}
\caption{
Optimized ${\rm S}_0$, ${\rm S}_1$, and ${\rm T}_1$ structures of rubrene obtained with RKS, ssUKS, and UKS, respectively. Selected bond lengths and dihedral angles are indicated. The ${\rm S}_1$ and ${\rm T}_1$ states are dominated by HOMO$\rightarrow$LUMO excitation character, with weights of 0.94 and 0.90, respectively. The figure also summarizes the corresponding vertical, adiabatic, and $0$--$0$ excitation energies relevant to the evaluation of ${\rm S}_1-2{\rm T}_1$. Here, $E_{{\rm S}_0}$ denotes the ground-state zero-point vibrational energy correction.
}
    \label{fig:myfigure}
\end{figure*}

As shown in Fig.~\ref{fig:myfigure}, the lowest singlet and triplet excitations retain predominantly single excitation character, with leading HOMO$\rightarrow$LUMO weights of 0.94 for ${\rm S}_1$ and 0.90 for ${\rm T}_1$. 
This supports the use of orbital-optimized single-reference approaches for the excited-state geometries. 
The optimized structures also show clear relaxation relative to ${\rm S}_0$, including changes in the central backbone bond lengths and torsional angles, which are expected to affect the adiabatic excitation energies.

\begin{table}[!hpbt]
\centering
\caption{
ADC(2)/cc-pVDZ excitation energies (in eV) of rubrene obtained with the $0$--$0$ protocol of Scheme-D, the corresponding adiabatic energies without ZPVE corrections, and vertical excitation energies evaluated at the ${\rm S}_0$, ${\rm S}_1$, and ${\rm T}_1$ geometries. The table also lists the derived ${\rm S}_1-{\rm T}_1$ and ${\rm S}_1-2{\rm T}_1$ energy differences together with the experimental reference values.
}
\label{tab:rubrene}
\begin{tabular}{lcccc}
\hline
Method & ${\rm S}_1$  & ${\rm T}_1$   & ${\rm S}_1-{\rm T}_1$   & ${\rm S}_1-2{\rm T}_1$   \\ 
\hline
  Scheme-D ($0$--$0$) &     2.17 & 1.05  &  1.12 &  0.07  \\
   Adiabatic  & 2.23 & 1.13 & 1.09 & $-$0.03     \\
    Scheme-G (vertical, $R_{{\rm T}_1})$  & 2.13 & 1.00 & 1.13  &  0.13  \\
     Scheme-H (vertical, $R_{{\rm S}_1})$ & 2.16 & 1.09 & 1.07 & $-$0.02  \\
     Scheme-I (vertical, $R_{{\rm S}_0})$ & 2.72       &  1.71   &    1.01    &   $-$0.70 \\
Experimental \cite{tsunoyama2017anion} & 2.42  &1.16  & 1.26 & 0.10   \\
\hline 
\end{tabular}
\end{table}

The excitation energies collected in Table~\ref{tab:rubrene} show that the same qualitative trends found for the benchmark set are retained for rubrene. 
Vertical excitation energies evaluated at the ${\rm S}_0$ geometry (Scheme~I) substantially overestimate both ${\rm S}_1$ and ${\rm T}_1$, with errors of about 0.3 to 0.6 eV relative to experiment, and they predict a strongly negative value of ${\rm S}_1-2{\rm T}_1$. 
Inclusion of excited-state relaxation lowers both excitation energies markedly. In this respect, Scheme~G already provides a reasonable vertical estimate of the singlet-fission metric, giving ${\rm S}_1-2{\rm T}_1 = 0.13$ eV, close to the experimental value of 0.10 eV. The fully relaxed $0$--$0$ values obtained from ssUKS and UKS geometries, 2.17 eV for ${\rm S}_1$ and 1.05 eV for ${\rm T}_1$, give ${\rm S}_1-2{\rm T}_1 = 0.07$ eV. By comparison, the corresponding adiabatic value without ZPVE correction is slightly negative, showing that rubrene lies very close to the energetic threshold and that even modest ZPVE contributions can change the sign of ${\rm S}_1-2{\rm T}_1$.

Overall, these results show that ssUKS and UKS-based excited-state geometries remain practically useful for larger molecules with predominantly single-excitation character. Although ADC(2)/cc-pVDZ still underestimates the absolute excitation energies, the relaxed state energetics move in the physically correct direction and recover the near-threshold singlet-fission energetics of rubrene with reasonable accuracy.

\section{Conclusions\label{sec:conclusion}}
Accurate prediction of adiabatic $0$--$0$ excitation energies requires reliable excited-state geometries in addition to an appropriate treatment of the electronic energies. Here, we assessed how LR-TDA, state-specific orbital-optimized UKS for ${\rm S}_1$, and UKS for ${\rm T}_1$ geometries propagate into ADC(2)-based adiabatic excitation energies and singlet--triplet gaps for a benchmark set spanning rigid aromatic systems, an inverted-gap chromophore, an anti-Kasha molecule, and a flexible carbonyl compound.

The most accurate results were obtained for Scheme-A, in which both ${\rm S}_1$ and ${\rm T}_1$ geometries were optimized with LR-TDA, yielding a mean error below 0.1 eV for the ADC(2)-based $0$--$0$ energies. Replacing these geometries with ssUKS for ${\rm S}_1$ and UKS for ${\rm T}_1$ in Scheme-D yields a computationally economical protocol that retains the accuracy for the predominantly single-excitation states considered here. In particular, UKS provides a reliable low-cost description of ${\rm T}_1$ minima, while ssUKS yields similarly useful ${\rm S}_1$ geometries when the targeted excited singlet state is dominated by a single determinant. 
 The close agreement between ssUKS/UKS-based and LR-TDA-based adiabatic $0$--$0$ energies indicates that, for the systems and electronic states considered here, spin-contamination effects are not large enough to significantly alter the relevant gradients and local curvatures of the excited-state surfaces.
  This practical similarity in accuracy is significant because ssUKS and UKS geometry optimizations retain an SCF-type structure and are therefore computationally more attractive than LR-TDA excited-state optimizations for larger molecules.

Vertical excitation energies were found to be much more sensitive to the choice of geometry than the corresponding ${\rm S}_1-{\rm T}_1$ gaps, for which partial error cancellation improves robustness. Application to rubrene further showed that relaxed-state energetics from ssUKS and UKS geometries provide a much more realistic description of the singlet-fission criterion $E({\rm S}_1)-2E({\rm T}_1)$ than vertical ${\rm S}_0$-based estimates.

Overall, the present results support ssUKS- and UKS-based excited-state geometries as useful low-cost approximations for evaluating adiabatic photophysical energy differences in larger molecules with dominant single-excitation character, while also clarifying that electronically mixed singlet states remain outside the robust domain of the present single-determinant strategy.

\section{Supplementary Information}
Supplementary Information provides: 
tables of basis-set comparisons, ZPVE corrections, expectation values of the total spin operator, torsional-profile energies for pyruvic acid, and data for azulene; 
figures showing frontier MOs and basis-set comparisons;
optimized geometries for the S$_0$, S$_1$ and T$_1$ states of all systems.

\section{Data Availability}
The data that support the findings of this study are within the article and its supplementary material.

\section{Acknowledgments}
We acknowledge the support of the 
Department of Atomic Energy, Government
of India, under Project Identification No.~RTI~4007. 
All calculations have been performed using the Helios computer cluster, 
which is an integral part of the MolDis 
Big Data facility, 
TIFR Hyderabad \href{http://moldis.tifrh.res.in}{(http://moldis.tifrh.res.in)}.

\section{Author Declarations}

\subsection{Author contributions}
\noindent 
{\bf AB}: 
Conceptualization (equal); 
Analysis (main); 
Data collection (main); 
Writing (main).
{\bf AM}: 
Conceptualization (equal); 
Analysis (supporting); 
Data collection (supporting); 
Writing (supporting).
{\bf RR}: Conceptualization (equal); 
Analysis (supporting); 
Data collection (supporting); 
Funding acquisition; 
Project administration and supervision; 
Resources; 
Writing (main).

\subsection{Conflicts of Interest}
The authors have no conflicts of interest to disclose.

\section{References}
\bibliography{ref} 
\end{document}



% ...





\clearpage

\singlespacing
\subsection*{Table of Contents}
\begin{enumerate}
    \item {\bf Table S1} Adiabatic $0$–$0$ excitation energies for S$_1$ and T$_1$ state for azulene, 5AP, pyruvic acid, anthracene, and fluoranthene computed using ADC(2) with different basis sets. (p.~S3)
    \item {\bf Table S2} Adiabatic 0–0 singlet–triplet energy gaps (S$_1$–T$_1$,)  for azulene, 5AP, pyruvic acid, anthracene, and fluoranthene computed using ADC(2) with different basis sets. (p.~S3)
    \item {\bf Table S3} Absolute electronic energies of pyruvic acid as a function of the ${\rm CH_3}$ torsional angle from $0^\circ$ to $60^\circ$.(p.~S4)
    \item {\bf Table S4} Zero-point vibrational energy corrections for the benchmark systems and rubrene at the optimized ${\rm S}_0$, ${\rm S}_1$, and ${\rm T}_1$ geometries. (p.~S4) 
    \item {\bf Table S5} Expectation values of $\langle {\hat S}^2\rangle$ obtained for the ssUKS description of the ${\rm S}_1$ state (p.~S5)
    \item {\bf Table S6} ADC(2)/cc-pVDZ ${\rm S}_2$, ${\rm T}_1$, and ${\rm S}_2-2{\rm T}_1$ energies of azulene obtained from selected $0$--$0$ and vertical schemes. (p. ~S5)
    \item {\bf Figure S1} Frontier MOs of Azulene and their energies from RKS and ssUKS DFT calculations. (p.~S6)
    \item {\bf Figure S2} Frontier MOs of Anthracene and their energies from RKS and ssUKS DFT calculations. (p.~S6)
    \item {\bf Figure S3} Frontier MOs of Fluoranthene and their energies from RKS and ssUKS DFT calculations. (p.~S7)
    \item {\bf Figure S4} Frontier MOs of Pyruvic acid and their energies from RKS and ssUKS DFT calculations. (p.~S7)
    \item {\bf Figure S5} Basis-set dependence of ADC(2) $0$--$0$ ${\rm S}_1$ excitation energies for 5AP, azulene, anthracene, fluoranthene, and pyruvic acid using the cc-pVDZ and cc-pVTZ basis sets. (p.~S8)
    \item {\bf Figure S6} Basis-set dependence of ADC(2) $0$--$0$ ${\rm T}_1$ excitation energies for 5AP, azulene, anthracene, fluoranthene, and pyruvic acid using the cc-pVDZ and cc-pVTZ basis sets. (p.~S9)
    \item {\bf Figure S7} Basis-set dependence of ADC(2) $0$--$0$ ${\rm S}_1-{\rm T}_1$ energy gap for 5AP, azulene, anthracene, fluoranthene, and pyruvic acid using the cc-pVDZ and cc-pVTZ basis sets. (p.~S10)
    \item Optimized Cartesian coordinates of
    5AP, azulene, anthracene, fluoranthene, pyruvic acid, and rubrene
    in S$_0$, S$_1$ and T$_1$ states (p.~S11-S41)
\end{enumerate}

\clearpage

\clearpage 

\begin{table*}[!htpb]
\centering
\setlength{\tabcolsep}{1.45pt}
\renewcommand{\arraystretch}{1.0}
\caption{
Adiabatic $0$–$0$ excitation energies for S$_1$ and T$_1$ state (in eV) for azulene, 5AP, pyruvic acid, anthracene, and fluoranthene computed using ADC(2) with different basis sets. 
Ground-state (${\rm S}_0$) geometries and zero-point vibrational energies were obtained at the RKS–$\omega$B97X-D3/def2-TZVP level, while excited-state (${\rm S}_1$ and ${\rm T}_1$) geometries and zero-point corrections were obtained using LR-TDA at the same level. Experimental reference values are shown for comparison, and mean signed error (MSE), mean absolute error (MAE), and standard deviation of the error (SDE) are reported. 
The basis sets cc-pVXZ and aug-cc-pVXZ are denoted here as VXZ and AVXZ, respectively.
}
\label{tab:00values}
\renewcommand{\arraystretch}{1.0}
\begin{tabular}{l ccc ccc ccc ccc ccc}
\hline
System & 
\multicolumn{2}{c}{ADC(2)/VDZ} &
&\multicolumn{2}{c}{ADC(2)/VTZ} &
&\multicolumn{2}{c}{ADC(2)/AVDZ} &
&\multicolumn{2}{c}{ADC(2)/AVTZ} &
&\multicolumn{2}{c}{Exp.} \\

\cline{2-3} \cline{5-6} \cline{8-9} \cline{11-12} \cline{14-15}

& S$_1$ & T$_1$
& & S$_1$ & T$_1$
& & S$_1$ & T$_1$
& & S$_1$ & T$_1$
& & S$_1$ & T$_1$ \\
\hline
5AP &$1.94$ &$2.08$ & & $1.99$ & $2.12$ & & $1.95$ & $2.07$  & & $2.00$ & $2.12$  & & $1.957$ & $2.003$ \\
Azulene &$1.84$ & $1.80$ & & $1.87$ & $1.82$  & & $1.82$ &  $1.76$ & & $1.86$ & $1.81$  & & $1.771$  & $1.722$  \\
Anthracene  &$3.39$ & $1.86$  & & $3.31$ & $1.90$ & & $3.25$ &  $1.83$ & & $3.26$ &$1.89$ & &  $3.433$ &  $1.872$ \\
Fluoranthene  &$3.20$ & $2.36$ & &  $3.18$ & $2.40$ & & $3.08$ & $2.31$  & & $3.14$ & $2.38$ & & $3.126$ & $2.321$  \\
Pyruvic acid  &$3.15$ & $2.75$  & & $3.13$ & $2.77$  & & $3.02$ &  $2.65$  & & $3.08$ & $2.73$& & $3.25$ & $2.962$ \\
MSE & $0.00$ & $0.01$  & & $0.01$ & $-0.03$  & & $0.08$ & $0.05$  & &  $0.04$ & $-0.01$ & & &  \\
MAE & $0.06$ & $0.08$  & & $0.09$ & $0.10$  & & $0.10$ & $0.09$  & & $0.10$ & $0.10$  & & & \\
SDE & $0.07$ & $0.11$  & & $0.09$ & $0.11$ & & $0.11$ & $0.14$  & & $0.11$ & $0.13$ & & &  \\
\hline
\end{tabular}
\end{table*}

\vspace{2cm}

\begin{table*}[!htpb]
\centering
\setlength{\tabcolsep}{1.25pt}
\renewcommand{\arraystretch}{1.0}
\caption{
Adiabatic 0–0 singlet–triplet energy gaps (S$_1$–T$_1$, in eV) (in eV) for azulene, 5AP, pyruvic acid, anthracene, and fluoranthene computed using ADC(2) with different basis sets. Ground-state (${\rm S}_0$) geometries and zero-point vibrational energies were obtained at the RKS–$\omega$B97X-D3/def2-TZVP level, while excited-state (${\rm S}_1$ and ${\rm T}_1$) geometries and zero-point corrections were obtained using LR-TDA at the same level. Experimental reference values are shown for comparison, and mean signed error (MSE), mean absolute error (MAE), and standard deviation of the error (SDE) are reported.
The basis sets cc-pVXZ and aug-cc-pVXZ are denoted here as VXZ and AVXZ, respectively.
}
\label{tab:00values}
\renewcommand{\arraystretch}{1.0}
\begin{tabular}{l c c c c c}
\hline
System & 
ADC(2)/VDZ &
ADC(2)/VTZ &
ADC(2)/AVDZ &
ADC(2)/AVTZ &
Exp. \\

\cline{2-2} \cline{3-3} \cline{4-4} \cline{5-5} \cline{6-6}

& S$_1$--T$_1$ 
& S$_1$--T$_1$
& S$_1$--T$_1$
& S$_1$--T$_1$
& S$_1$--T$_1$ \\

\hline
5AP & $-0.14$&   $-0.12$ &   $-0.12$ &  $-0.12$ &   $-0.047$\\
Azulene & $0.04$ &  $0.05$ &  $0.05$ &   $0.05$ &   $0.049$ \\
Anthracene  &  $1.53$ &  $1.41$ &   $1.41$ &  $1.38$ &   $1.561$ \\
Fluoranthene  & $0.84$ &   $0.78$ &  $0.77$ &   $0.76$ &   $0.805$ \\
Pyruvic acid  & $0.40$ &  $0.37$ &  $0.37$  &   $0.35$  &   $0.29$\\
MSE &  $0.00$ &   $0.03$ &   $0.04$ &   $0.05$ &   \\
MAE &  $0.06$ &   $0.07$ &  $0.07$ &  $0.07$ &   \\
SDE &  $0.07$  &   $0.08$ &  $0.08$ &   $0.08$ &   \\
\hline
\end{tabular}
\end{table*}

\clearpage

\begin{table*}
\centering
\caption{
Absolute electronic energies of pyruvic acid in Hartree as a function of the ${\rm CH_3}$ torsional angle from $0^\circ$ to $60^\circ$. Ground-state ${\rm S}_0$ energies were obtained with RKS, while the lowest excited singlet and triplet energies were obtained with LR-TDA, ssUKS, and UKS, as indicated. For the ssUKS ${\rm S}_1$ values, the corresponding $\langle {\hat S}^2\rangle$ expectation values are given in parentheses. The tabulated values provide the numerical data underlying the torsional energy profiles discussed in the main text and show that the spin contamination of the ssUKS ${\rm S}_1$ solution remains nearly constant along the torsional coordinate.
}
\begin{tabular}{c ccccc}
\hline
\multirow{2}{*}{Torsion angle} & \multicolumn{5}{l}{Absolute electronic energy} \\
\cline{2-6}
 & ${\rm S}_0$ (RKS) & ${\rm S}_1$ (LR-TDA) & ${\rm T}_1$ (LR-TDA) & ${\rm S}_1$ (ssUKS) & ${\rm T}_1$ (UKS) \\
\hline
$0^\circ$  & $-$9318.537 & $-$9314.884 & $-$9315.455 & $-$9315.182 (1.011) & $-$9315.380 \\
$10^\circ$ & $-$9318.534 & $-$9314.888 & $-$9315.458 & $-$9315.185 (1.011) & $-$9315.383 \\
$20^\circ$ & $-$9318.527 & $-$9314.897 & $-$9315.465 & $-$9315.195 (1.011) & $-$9315.392 \\
$30^\circ$ & $-$9318.518 & $-$9314.911 & $-$9315.476 & $-$9315.209 (1.012) & $-$9315.404 \\
$40^\circ$ & $-$9318.509 & $-$9314.924 & $-$9315.487 & $-$9315.222 (1.012) & $-$9315.416 \\
$50^\circ$ & $-$9318.503 & $-$9314.934 & $-$9315.494 & $-$9315.232 (1.012) & $-$9315.425 \\
$60^\circ$ & $-$9318.501 & $-$9314.939 & $-$9315.498 & $-$9315.236 (1.012) & $-$9315.429 \\
\hline
\end{tabular}
\label{tab:geom_benchmark}
\end{table*}
\vspace{6cm}

\begin{table*}[h!]
\centering
\caption{
Zero-point vibrational energy corrections (in eV) for the benchmark systems and rubrene at the optimized ${\rm S}_0$, ${\rm S}_1$, and ${\rm T}_1$ geometries considered in this work. Ground-state geometries were obtained with RKS, ${\rm S}_1$ geometries with LR-TDA or ssUKS, and ${\rm T}_1$ geometries with LR-TDA or UKS, as indicated in the table. All DFT geometry optimizations and frequency calculations were performed at the $\omega$B97X-D3/def2-TZVP level. No LR-TDA excited-state geometries were computed for rubrene.
}
\begin{tabular}{lcc ccc}
\hline
System & ${\rm S}_0$ (RKS) & ${\rm S}_1$ (LR-TDA) & ${\rm T}_1$ (LR-TDA) & ${\rm S}_1$ (ssUKS) & ${\rm T}_1$ (UKS) \\ 
\hline
 5AP          & $3.556$ & $3.508$ & $3.463$ & $3.463$ & $3.407$  \\
 Azulene     & $3.995$ & $3.984$ & $3.894$ & $3.950$ & $3.888$ \\
 Anthracene   & $5.325$ & $5.265$ & $5.215$ & $5.239$  & $5.212$  \\
 Fluoranthene & $5.674$ & $5.605$ & $5.533$ & $5.580$ & $5.545$  \\
 Pyruvic Acid & $1.965$ & $1.884$ & $1.892$ & $1.876$ & $1.879$  \\
 Rubrene      &$15.475$ & - & - & $15.423$ & $15.397$  \\
\hline 
\end{tabular}%
\label{tab:ZPVE}
\end{table*}

\newpage
\begin{table}[h!]
\centering
\caption{
Expectation values of $\langle {\hat S}^2\rangle$ obtained for the ssUKS description of the ${\rm S}_1$ state at the corresponding optimized ${\rm S}_1$ geometries of the benchmark systems.
}
\label{tab:S2_values}
\begin{tabular}{ll}
\hline
{Systems} & \textbf {$\langle S^2 \rangle$} \\
\hline
5AP  &  1.172  \\
Azulene  &  1.093  \\
Anthracene & 1.029 \\
Fluoranthene  & 1.057    \\
Pyruvic Acid  & 1.012  \\
\hline
\end{tabular}
\end{table}

\vspace{10cm}

\begin{table*}[!htpb]
\centering
\caption{
ADC(2)/cc-pVDZ ${\rm S}_2$, ${\rm T}_1$, and ${\rm S}_2-2{\rm T}_1$ energies of azulene, in eV, obtained from selected $0$--$0$ and vertical schemes using different combinations of ground- and excited-state geometry optimization protocols. The ground-state ${\rm S}_0$ geometry was optimized with RKS, while excited-state geometries were optimized with LR-TDA, ssUKS, or UKS, as indicated. For the Scheme-D results, two alternative ssUKS occupation patterns were considered for the ${\rm S}_2$ state, corresponding to ${\rm HOMO}\rightarrow{\rm LUMO}+1$ and  ${\rm HOMO}-1\rightarrow{\rm LUMO}$. 
}
\label{tab:geomeffect}
\small
\begin{tabular}{l l ccc c ccc}
\hline
Scheme & Excitation type & \multicolumn{4}{l}{Geometry optimization } & \multicolumn{3}{l}{Energies} \\
\cline{3-5} \cline{7-9}
      &          & S$_0$ &  S$_2$ & T$_1$ && S$_2$ & T$_1$ & S$_2$-2T$_1$ \\
\hline
A     & 0-0      & RKS & LR-TDA  & LR-TDA && 3.719 & 1.802 & 0.115 \\
D$_1$ (HOMO$\rightarrow$LUMO$+1$)   & 0-0      & RKS &  ssUKS & UKS && 3.870 & 1.777 & 0.316 \\
D$_2$ (HOMO$-1\rightarrow$LUMO)    & 0-0      & RKS &  ssUKS & UKS && 3.873 & 1.777 & 0.319 \\
G     & Vertical & --   & -- & UKS && 3.565 & 1.467 & 0.631  \\
I     & Vertical & RKS  & -- &  - && 3.980 & 2.307 &$-$0.634    \\
\hline
\end{tabular}
\end{table*}

\clearpage

\begin{figure*}[!htbp]
    \centering
    \includegraphics[width=\textwidth]{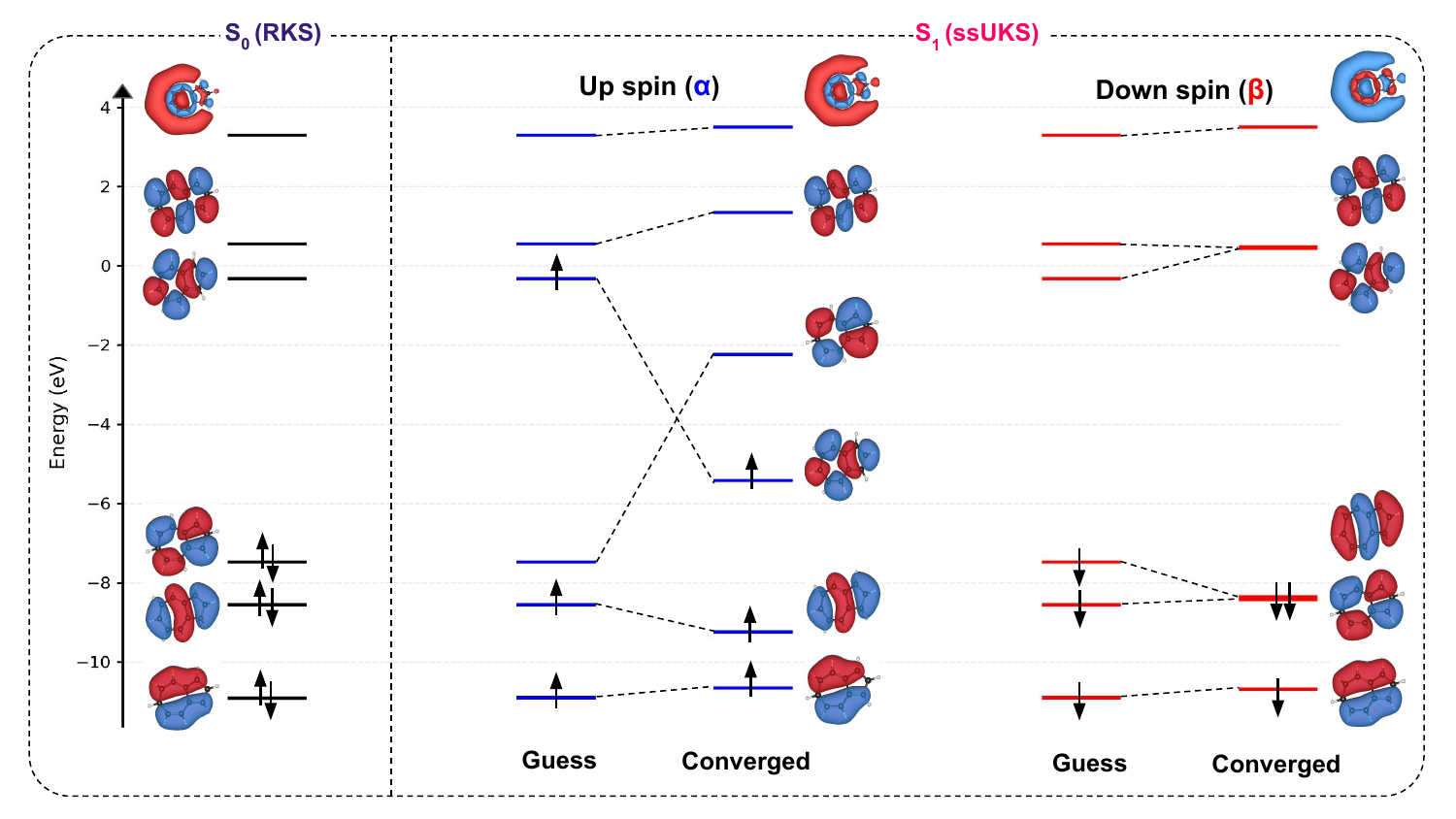}
    \caption{
    Frontier MOs of Azulene and their energies from RKS and ssUKS DFT calculations. 
    MOs from  ssUKS calculation are shown separately for the $\alpha$ and $\beta$ spin channels.
    }
    \label{fig:AzulenessUKSMOs}
\end{figure*}

\begin{figure*}[!htbp]
    \centering
    \includegraphics[width=\textwidth]{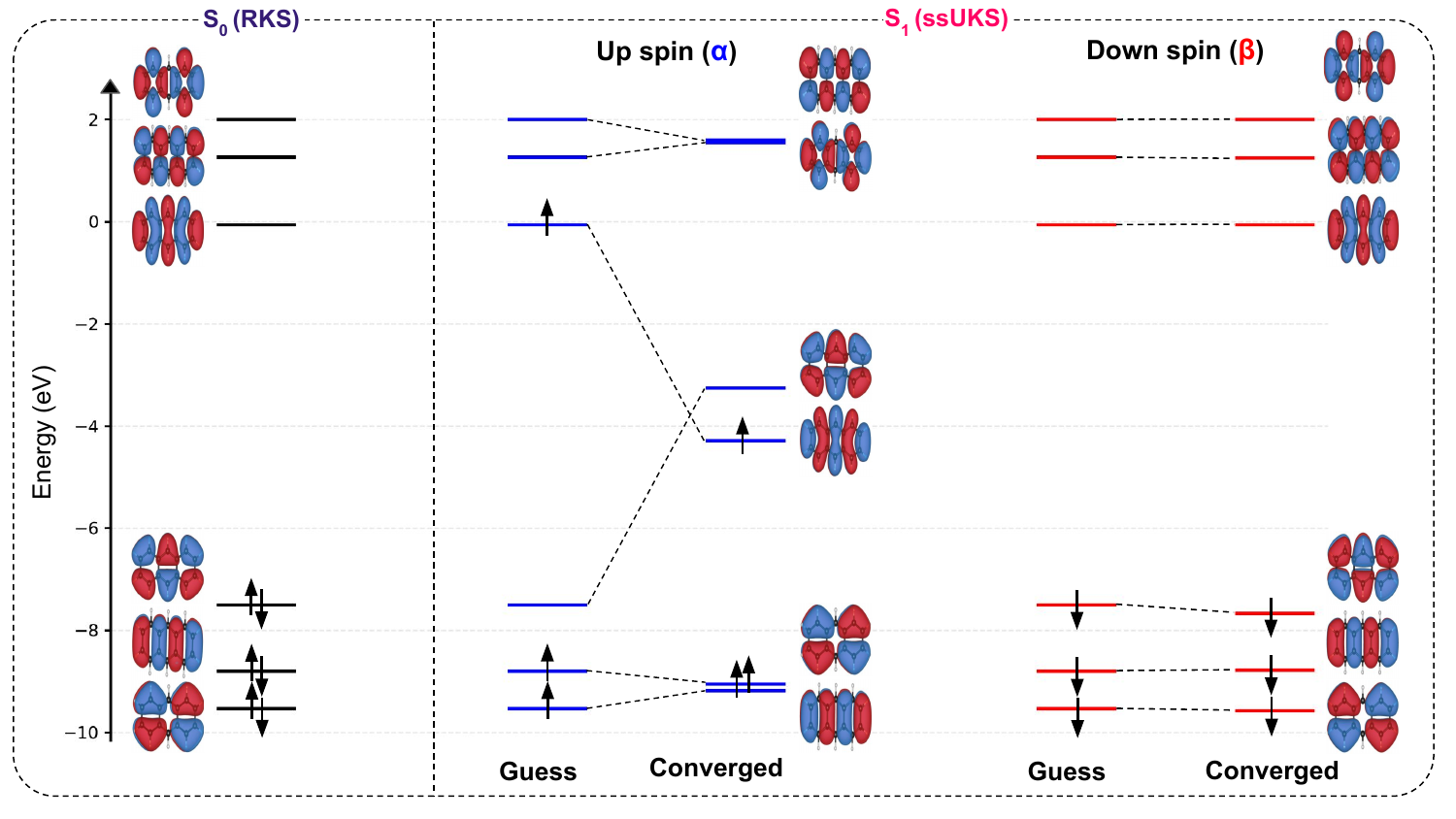}
    \caption{
    Frontier MOs of Anthracene and their energies from RKS and ssUKS DFT calculations. 
 MOs from  ssUKS calculation are shown separately for the $\alpha$ and $\beta$ spin channels.
    }
    \label{fig:Anthracene_ssUKSMOs}
\end{figure*}

\begin{figure*}[!htbp]
    \centering
    \includegraphics[width=\textwidth]{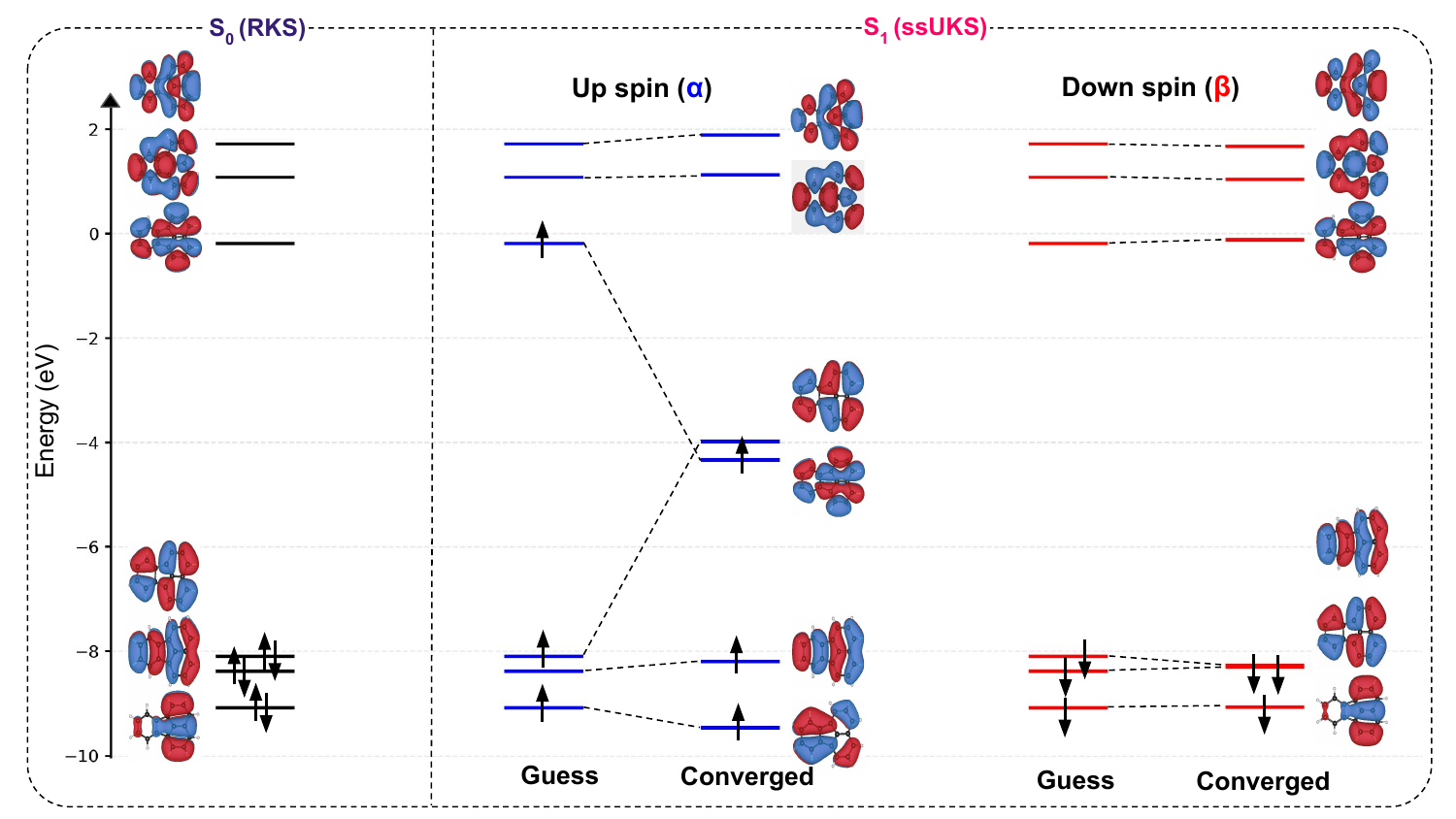}
    \caption{
    Frontier MOs of Fluoranthene and their energies from RKS and ssUKS DFT calculations. 
 MOs from  ssUKS calculation are shown separately for the $\alpha$ and $\beta$ spin channels.
    }
    \label{fig:FluoranthenessUKSMOs}
\end{figure*}

\begin{figure*}[!htbp]
        \centering
        \includegraphics[width=\linewidth]{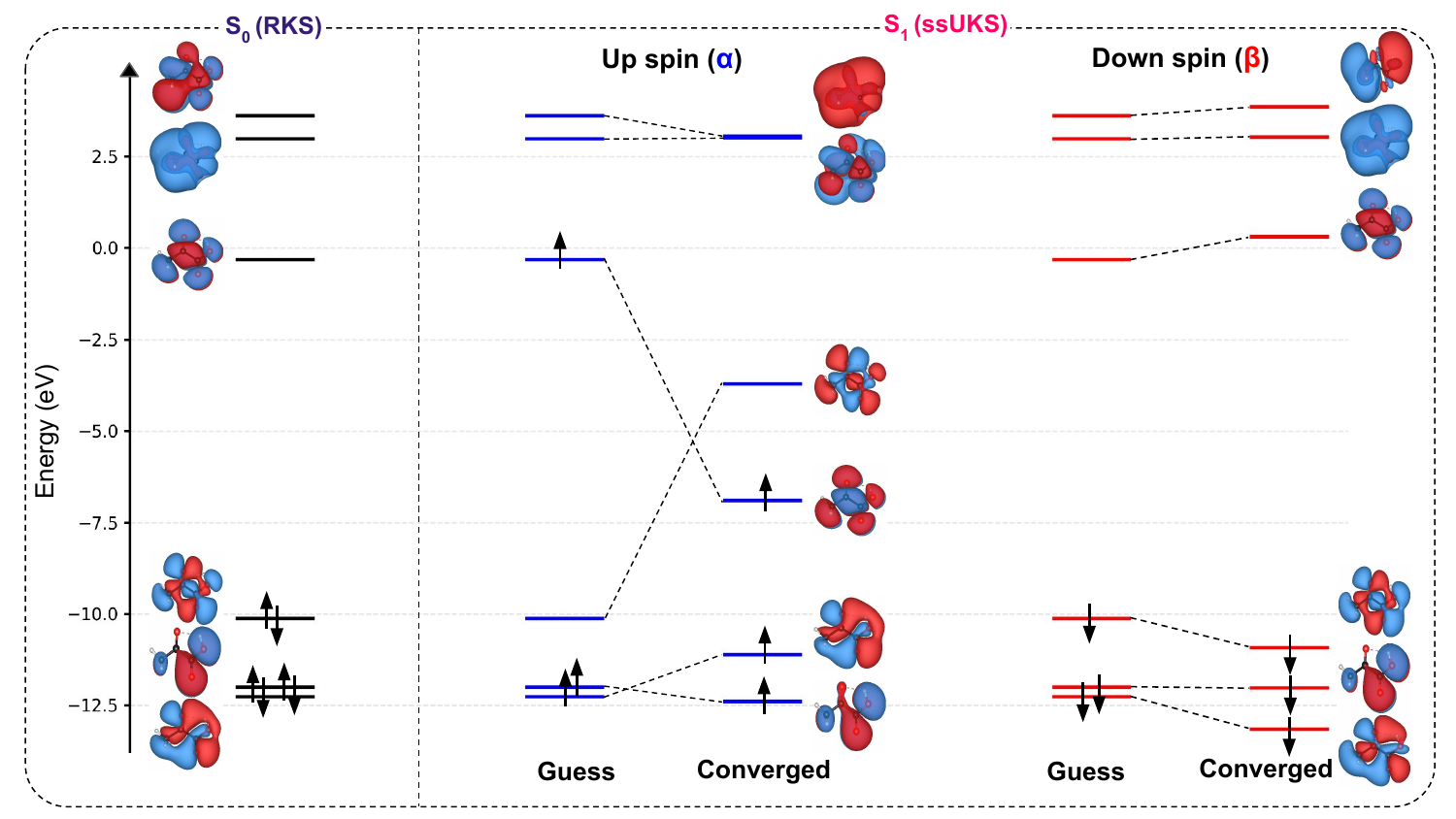}
        \caption{
    Frontier MOs of Pyruvic acid and their energies from RKS and ssUKS DFT calculations. 
 MOs from  ssUKS calculation are shown separately for the $\alpha$ and $\beta$ spin channels.
    }
    \label{fig:Pyruvic_acid_ssUKSMOs}
\end{figure*}

\begin{figure*}[!htbp]
        \centering
        \includegraphics[width=\linewidth]{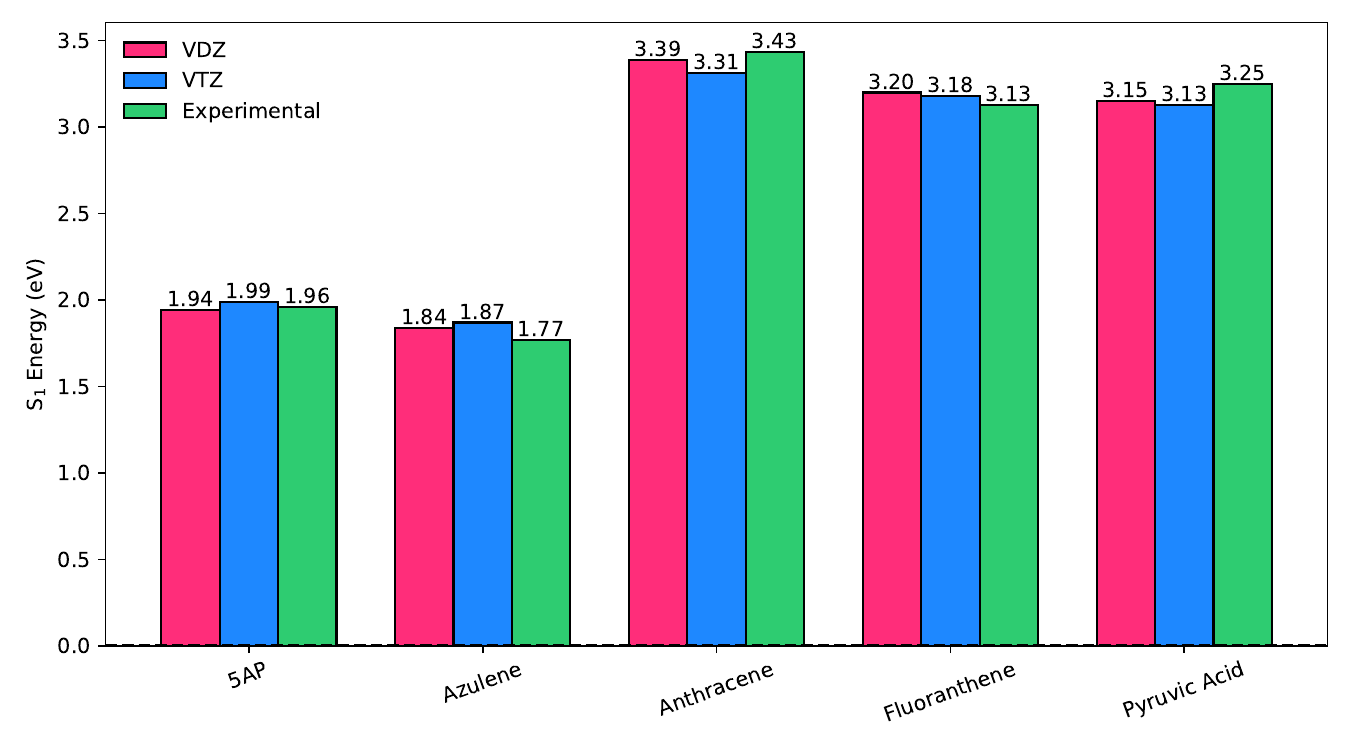}
        \includegraphics[width=\linewidth]{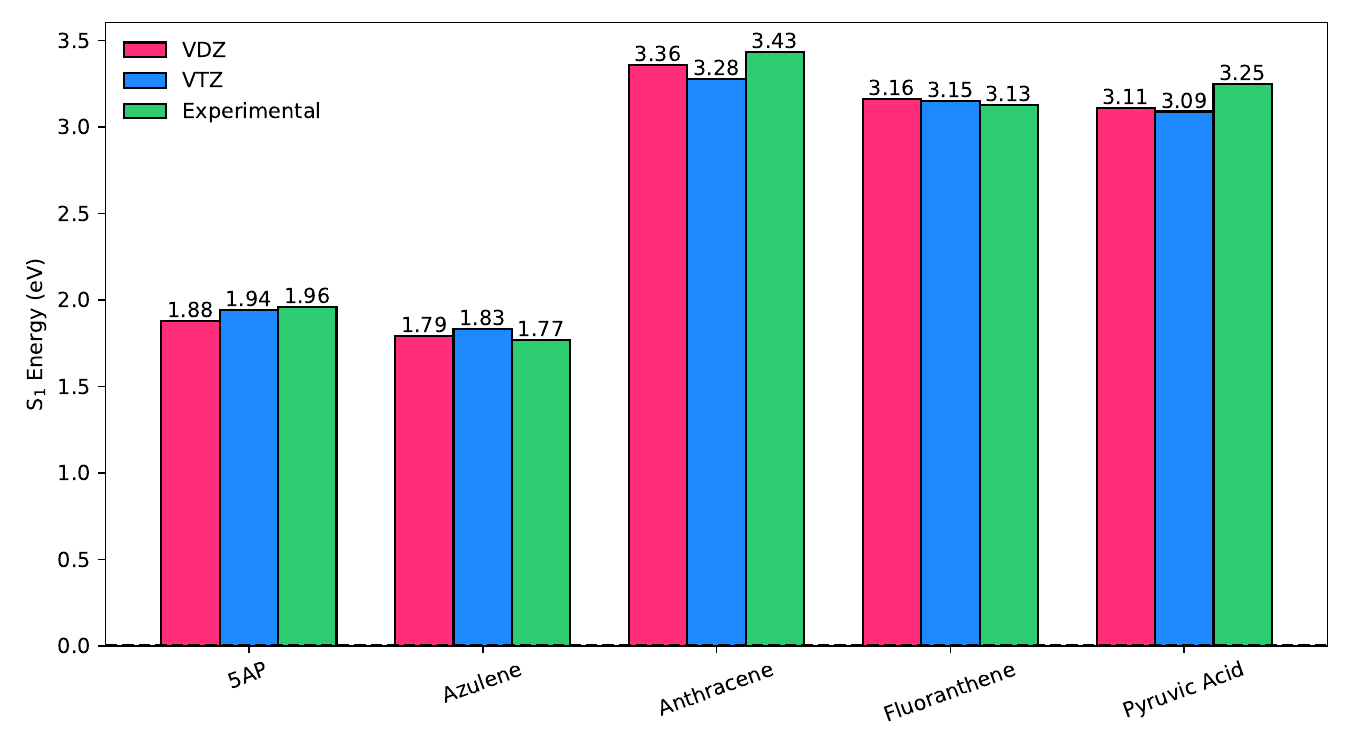}
        \caption{
Basis-set dependence of ADC(2) $0$--$0$ ${\rm S}_1$ excitation energies for 5AP, azulene, anthracene, fluoranthene, and pyruvic acid using the cc-pVDZ and cc-pVTZ basis sets. 
The \ul{top} panel corresponds to LR-TDA optimized ${\rm S}_1$ geometries and ZPVE corrections, and the \ul{bottom} panel to ssUKS-DFT optimized ${\rm S}_1$ geometries and ZPVE corrections. 
In both cases, the ${\rm S}_0$ geometry and ZPVE correction were obtained at the RKS $\omega$B97X-D3/def2-TZVP level.
    }
    \label{fig:S1_Scheme_AD}
\end{figure*}

\begin{figure*}[!htbp]
        \centering
        \includegraphics[width=\linewidth]{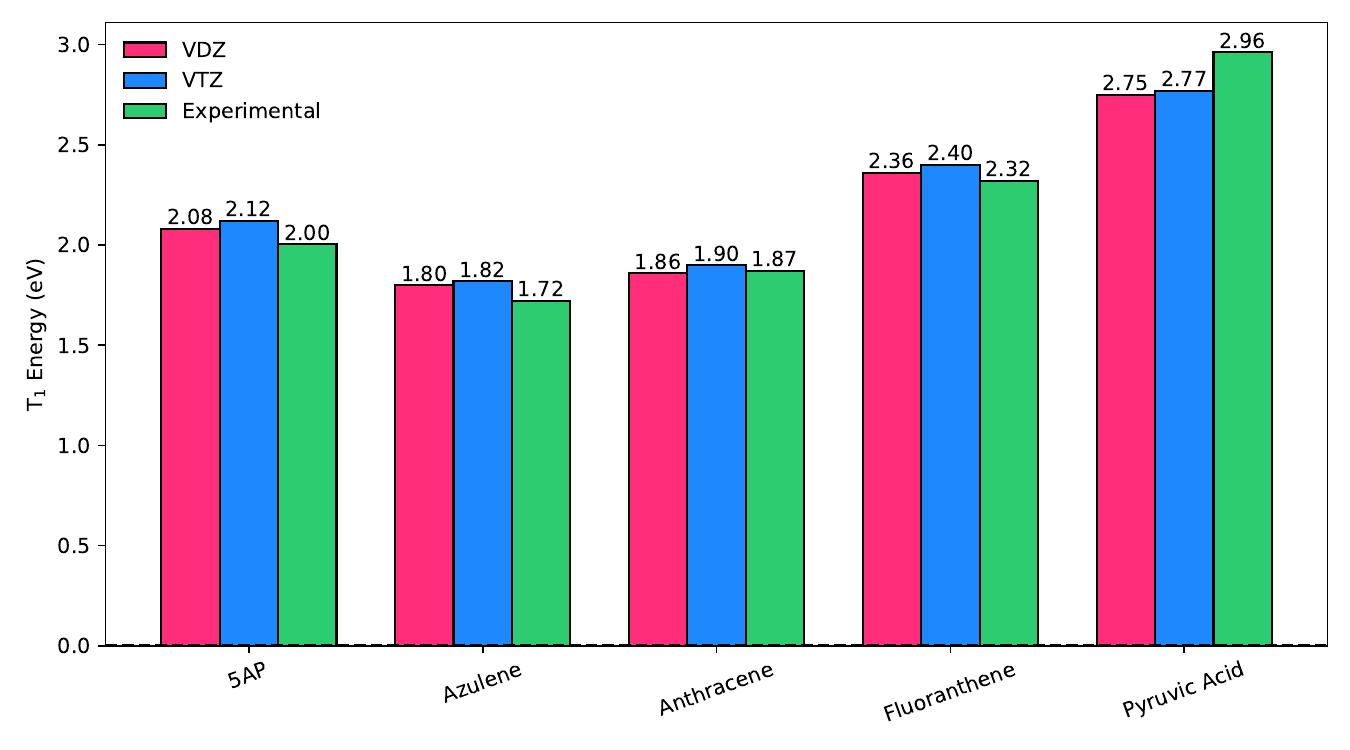}
        \includegraphics[width=\linewidth]{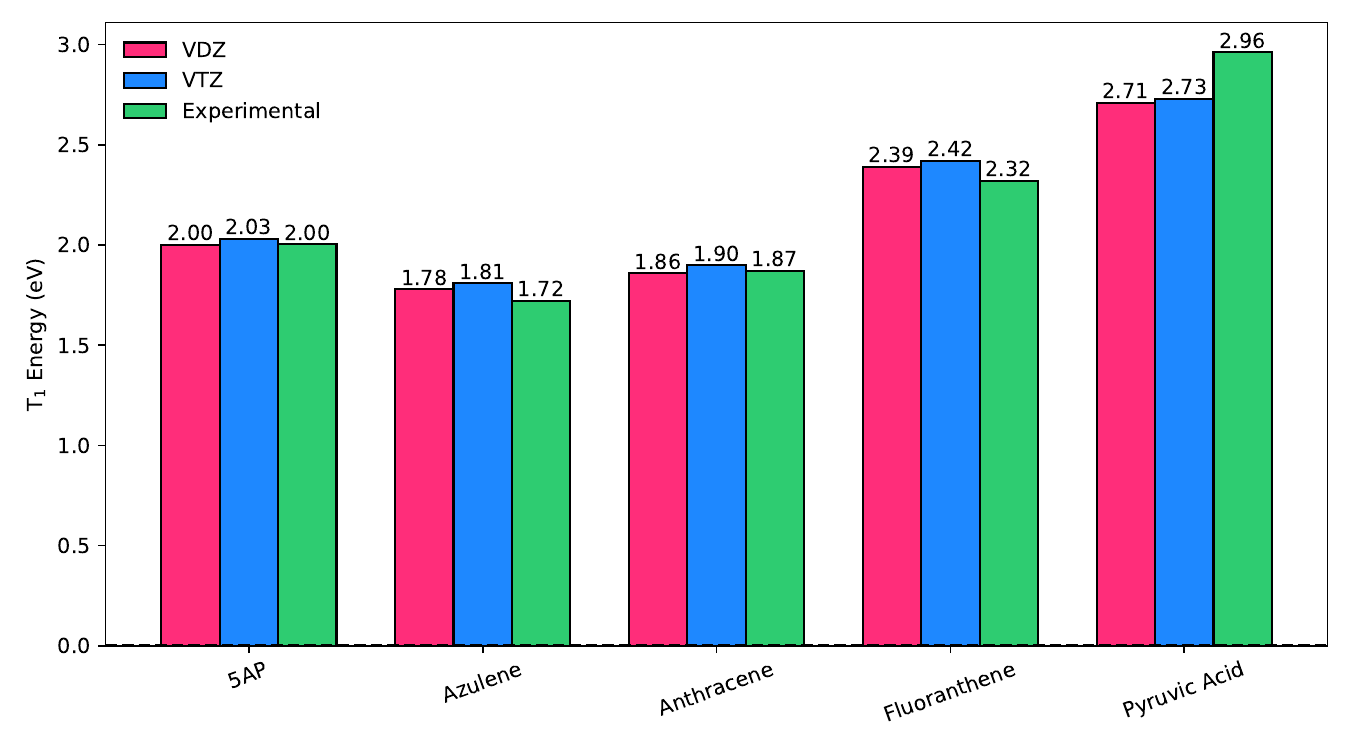}
    \caption{
    Basis-set dependence of ADC(2) $0$--$0$ ${\rm T}_1$ excitation energies for 5AP, azulene, anthracene, fluoranthene, and pyruvic acid using the cc-pVDZ and cc-pVTZ basis sets. The \ul{top} panel corresponds to LR-TDA optimized ${\rm T}_1$ geometries and ZPVE corrections, and the \ul{bottom} panel to UKS-DFT optimized ${\rm T}_1$ geometries and ZPVE corrections. In both cases, the ${\rm S}_0$ geometry and ZPVE correction were obtained at the RKS $\omega$B97X-D3/def2-TZVP level.
    }
    \label{fig:T1_Scheme_AD}
\end{figure*}

\begin{figure}
    \centering
    \includegraphics[width=\linewidth]{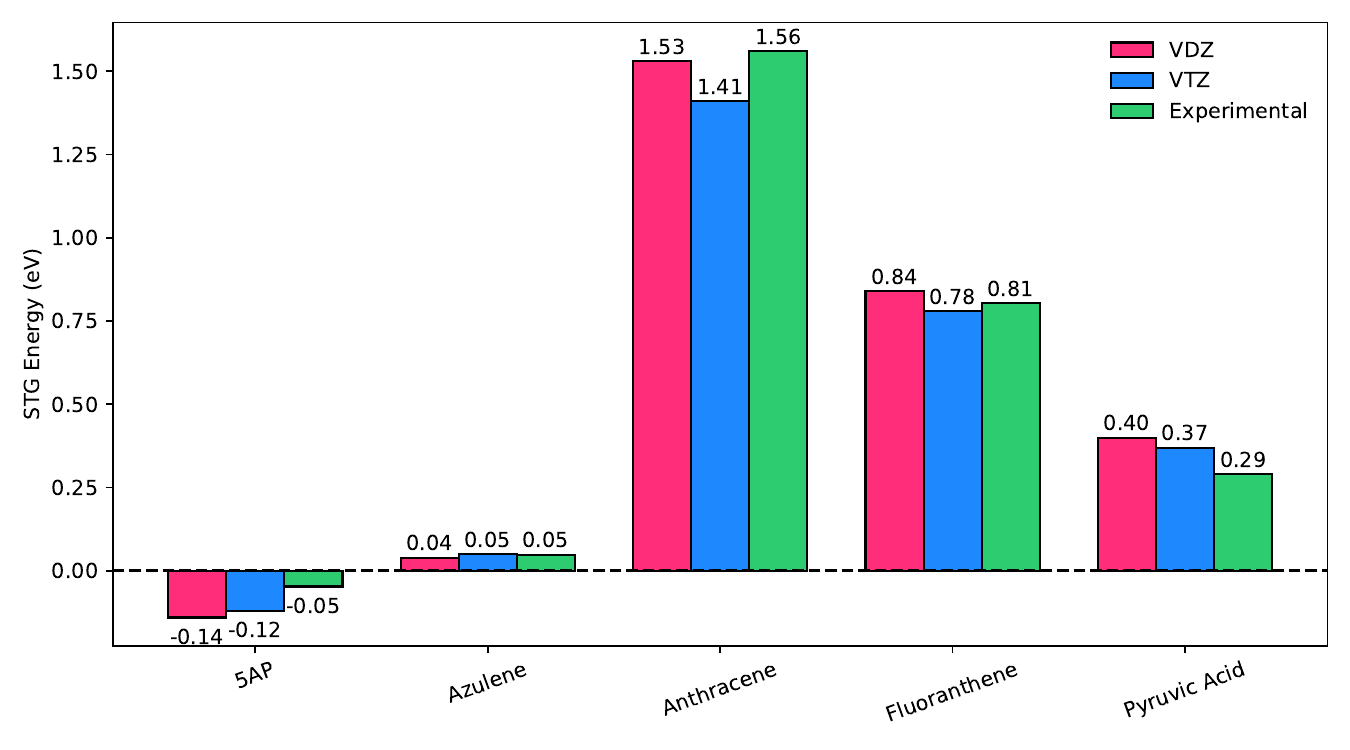}
    \includegraphics[width=\linewidth]{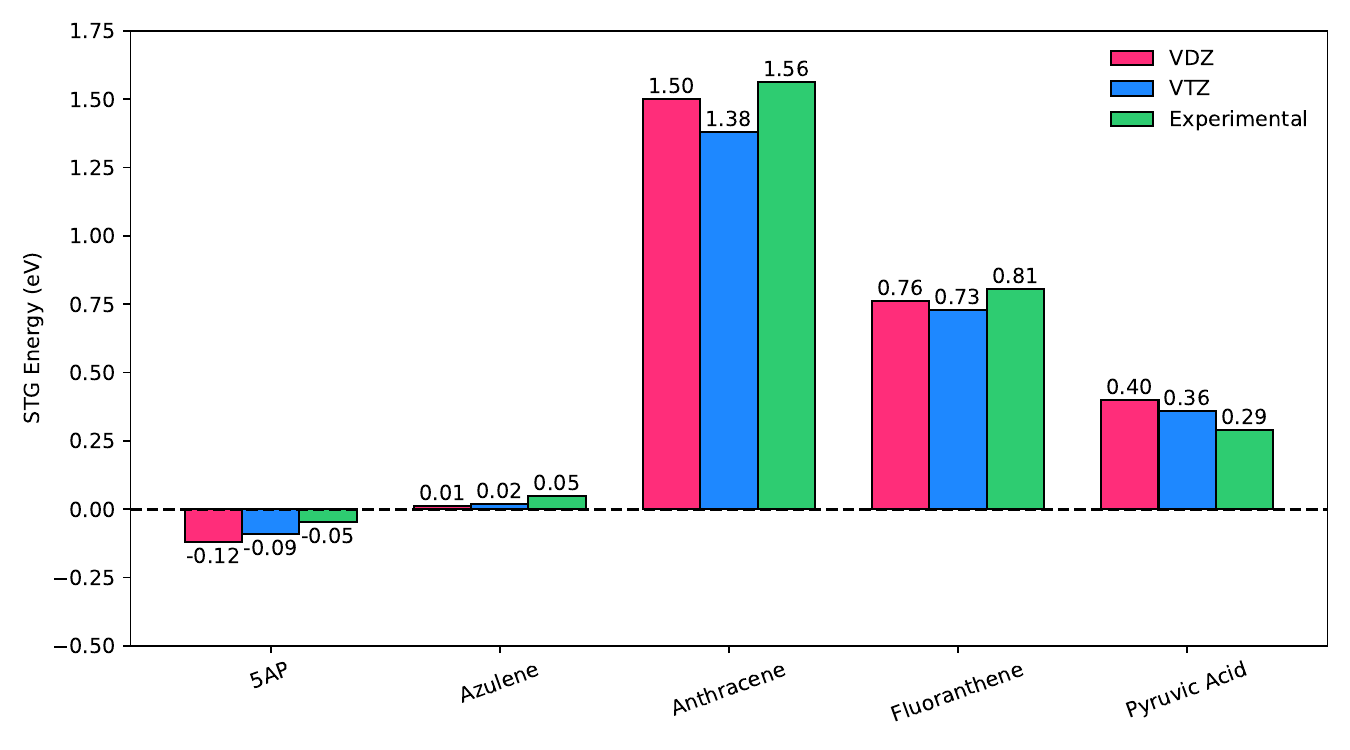}
        \caption{
Basis-set dependence of ADC(2) $0$--$0$ ${\rm S}_1-{\rm T}_1$ energy gap for 5AP, azulene, anthracene, fluoranthene, and pyruvic acid using the cc-pVDZ and cc-pVTZ basis sets. The \ul{top} panel corresponds to LR-TDA optimized excited state geometries and ZPVE corrections, and the \ul{bottom} panel to ssUKS/UKS-DFT optimized excited state geometries and ZPVE corrections. In both cases, the ${\rm S}_0$ geometry and ZPVE correction were obtained at the 
RKS $\omega$B97X-D3/def2-TZVP level.
        }
    \label{fig:STG_Scheme_AD}
\end{figure}

\clearpage


\singlespacing
\small
{ 
\begin{verbatim}
--------------------------------------------------------------------------
CARTESIAN COORDINATES (ANGSTROEM), RKS-wB97X-D3/def2-TZVP 
MOLECULE: 5AP, STATE: S0
--------------------------------------------------------------------------
18
5AP S0 RKS-wB97X-D3/def2-TZVP 
  N          -4.81601012595798      1.06801164203254      0.00764666086319
  C          -3.63511366093668      1.67122440436472      0.09437919833225
  C          -5.85422444235534     -2.26111694060306     -0.16288206786465
  C          -1.20770637769009     -0.91930365898399      0.13367254927477
  C          -4.81560428290298     -0.25464185431362     -0.03782286408779
  N          -2.44647319293188      1.12140470900551      0.14009532911017
  H          -6.79740686194422     -2.79651124174739     -0.23141072569696
  H          -0.30296448357964     -0.33451905868471      0.20073847837260
  C          -2.40828171655508     -0.22331672082657      0.09419578651119
  C          -1.20805901706011     -2.29866608065285      0.08545793746257
  H          -0.26579075565233     -2.83257685086583      0.11634927846857
  N          -4.77128127638387     -2.99833099174384     -0.12942017205922
  C          -2.38752261293015     -3.01006545182932     -0.00192058297893
  H          -3.66353062135250      2.75691815882521      0.13187714177999
  N          -3.59018242353214     -0.94906669898833      0.00528287572205
  N          -5.94697802632893     -0.93603964820245     -0.12305036936703
  H          -2.42185779670602     -4.08815045001941     -0.04161781599923
  C          -3.60152629670005     -2.33793267396661     -0.04366663734354
--------------------------------------------------------------------------
\end{verbatim}
}
\clearpage 

\singlespacing
\small
{ 
\begin{verbatim}
--------------------------------------------------------------------------
CARTESIAN COORDINATES (ANGSTROEM), LR-TDA-wB97X-D3/def2-TZVP 
MOLECULE: 5AP, STATE: S1
--------------------------------------------------------------------------
18
5AP S1 LR-TDA-wB97X-D3/def2-TZVP 
  N          -4.82199798372385      1.07334452213664     -0.01507342458502
  C          -3.65003272237269      1.68932770430589      0.09234706818209
  C          -5.87706805063713     -2.25701980365433     -0.17044623989135
  C          -1.20000860730838     -0.92243639473629      0.13129997821092
  C          -4.82959241266783     -0.24711216058498     -0.03215442773825
  N          -2.44256803678733      1.10198237011669      0.13630278380071
  H          -6.80890471368443     -2.80184321344883     -0.25235317526276
  H          -0.29284978873625     -0.33623954180798      0.16875595342487
  C          -2.40488199751203     -0.21108642116713      0.12030111129408
  C          -1.18867108726426     -2.30974778246092      0.08743855794264
  H          -0.24738018309247     -2.84253442175793      0.09803360704278
  N          -4.75291066378000     -2.99139400636804     -0.12850910182968
  C          -2.38104819218414     -3.01532557493344      0.00349535303367
  H          -3.66414978623312      2.77141438443714      0.12304184306844
  N          -3.60118072508007     -0.94916673463117      0.10106887249301
  N          -5.95314924847046     -0.93103312806374     -0.15015639996620
  H          -2.41211119772992     -4.09374851305294     -0.06093731979663
  C          -3.61200857423561     -2.35006069152863     -0.01455103892332
---------------------------------------------------------------------------
\end{verbatim}
}
\clearpage 

\singlespacing
\small
{ 
\begin{verbatim}
--------------------------------------------------------------------------
CARTESIAN COORDINATES (ANGSTROEM), ssUKS-wB97X-D3/def2-TZVP 
MOLECULE: 5AP, STATE: S1
--------------------------------------------------------------------------
18
5AP S1 ssUKS-wB97X-D3/def2-TZVP 
  N          -4.82608126350290      1.08012599068708      0.00688352643652
  C          -3.65938257693079      1.69423096408853      0.09324251789786
  C          -5.88639296800397     -2.25204200277302     -0.16426925532950
  C          -1.19321146444848     -0.92705101382857      0.13481514183678
  C          -4.83821346459720     -0.24182596244131     -0.03794634243036
  N          -2.43568677531872      1.09521998939654      0.13901415884284
  H          -6.81408098210698     -2.80486432719449     -0.23237065148868
  H          -0.28765423801838     -0.34094279321551      0.20090936818380
  C          -2.40903238608061     -0.20706819810099      0.09620345963681
  C          -1.18371405635324     -2.31253170105681      0.08659124274272
  H          -0.24266969879032     -2.84573125613671      0.11665468749393
  N          -4.74325832445592     -2.99401222047201     -0.12882669874351
  C          -2.37354317206370     -3.01858098591499     -0.00191874283215
  H          -3.66476845835889      2.77562965302758      0.13069025621544
  N          -3.59597436086063     -0.94608547027091      0.00992086809356
  N          -5.96247930551451     -0.93354508452853     -0.12410809081240
  H          -2.40852101336264     -4.09789313944780     -0.04364759990692
  C          -3.61584946273211     -2.34571184901807     -0.04393384533675
---------------------------------------------------------------------------
\end{verbatim}
}
\clearpage 

\singlespacing
\small
{ 
\begin{verbatim}
--------------------------------------------------------------------------
CARTESIAN COORDINATES (ANGSTROEM), LR-TDA-wB97X-D3/def2-TZVP 
MOLECULE: 5AP, STATE: T1
--------------------------------------------------------------------------
18
5AP T1 LR-TDA-wB97X-D3/def2-TZVP 
  N          -4.82898251829125      1.07221784799842      0.01050526397528
  C          -3.64600163445107      1.68279031964536      0.09487330211343
  C          -5.86966902087557     -2.25755616983484     -0.16565061194892
  C          -1.20304565645807     -0.92437503029024      0.13072587811835
  C          -4.82765375947311     -0.24803193204540     -0.03480761981256
  N          -2.44544602453741      1.10641401591639      0.13621396182472
  H          -6.80368712515746     -2.80158175701517     -0.24008842341456
  H          -0.30181132873976     -0.33033896399659      0.19515082411793
  C          -2.40256842440418     -0.21255883327830      0.09390054218632
  C          -1.17529368110009     -2.31736024873561      0.08762878209949
  H          -0.23647279130858     -2.85001449364817      0.11842615987799
  N          -4.75817022330287     -2.99118313711652     -0.12923152396105
  C          -2.38082114809809     -3.01128059784139      0.00061125640970
  H          -3.66231867136141      2.76557314837269      0.13241868780503
  N          -3.60950438689362     -0.93852270394579      0.00956557371820
  N          -5.95745580420077     -0.92715072343018     -0.12201499428887
  H          -2.42343903494942     -4.09107977842960     -0.03983088132309
  C          -3.60817273789723     -2.34864036952504     -0.04049217699741
---------------------------------------------------------------------------
\end{verbatim}
}
\clearpage 

\singlespacing
\small
{ 
\begin{verbatim}
--------------------------------------------------------------------------
CARTESIAN COORDINATES (ANGSTROEM), UKS-wB97X-D3/def2-TZVP 
MOLECULE: 5AP, STATE: T1
--------------------------------------------------------------------------
18
5AP T1 UKS-wB97X-D3/def2-TZVP 
  N          -4.81931394367852      1.08909944029043      0.00889980973470
  C          -3.67533264913060      1.69772631071413      0.09265592600177
  C          -5.89738191663706     -2.24003559650614     -0.16384935536527
  C          -1.18900402548745     -0.93463070611006      0.13498943423422
  C          -4.84057295759363     -0.24029295464353     -0.03812340956193
  N          -2.42810677869619      1.08495136733037      0.13946605716363
  H          -6.81798585351260     -2.80489371420287     -0.23073596548782
  H          -0.28639205745305     -0.34468163354131      0.20244103476406
  C          -2.41905754734373     -0.20782624930267      0.09456851772823
  C          -1.18097828228341     -2.31407658040815      0.08653265152466
  H          -0.23943887190299     -2.84763615109944      0.11752766475386
  N          -4.73060429784140     -2.99528621752340     -0.12873793688522
  C          -2.36508782502361     -3.01829113318037     -0.00226174393433
  H          -3.66685745184076      2.77909550693711      0.13106480805913
  N          -3.59213112618926     -0.94805572166513      0.00560250613842
  N          -5.96700424062747     -0.94438731177713     -0.12283122420832
  H          -2.40510883532079     -4.09714633652725     -0.04363264962920
  C          -3.62015531093745     -2.33631172598456     -0.04567212453058
---------------------------------------------------------------------------
\end{verbatim}
}
\clearpage

\singlespacing
\small
{ 
\begin{verbatim}
--------------------------------------------------------------------------
CARTESIAN COORDINATES (ANGSTROEM), RKS-wB97X-D3/def2-TZVP 
MOLECULE: Azulene, STATE: S0
--------------------------------------------------------------------------
18
Azulene S0 RKS-wB97X-D3/def2-TZVP 
  C          -2.22361839781327      1.99180869913452     -0.01193924755441
  C          -1.00677799419412      2.65869402435111      0.00447253091977
  C          -0.18011038787123     -0.39937830546468     -0.11721456966792
  H           2.65838535020005     -0.23513153405514     -0.13062945274495
  C          -2.46162890419260      0.62479308634235     -0.06231484772597
  C           0.27412056646277      2.13451470528793     -0.02536964975186
  C           2.53327772233351      1.97452766737111     -0.04823886222617
  C           2.05901256223643      0.66317538212784     -0.09373974074762
  C           0.66222619659525      0.69878796753091     -0.08201254562784
  H           1.52924631267267      3.94640739152521      0.03388038104782
  C           1.46302696572560      2.86869205100361     -0.00672000529917
  H           0.31905281603374     -1.36415308887988     -0.15738721756860
  H           3.57775044775206      2.25666971951168     -0.04543316525619
  H          -1.06220788798049      3.74345851534346      0.04660399215652
  C          -1.56719023170951     -0.43636305335770     -0.10883269163351
  H          -3.50985381583974      0.34150849643544     -0.06580504591269
  H          -3.10670400994178      2.62058802693227      0.01772712886954
  H          -2.01302821736934     -1.42431275804005     -0.14353237517676
--------------------------------------------------------------------------
\end{verbatim}
}
\clearpage

\singlespacing
\small
{ 
\begin{verbatim}
--------------------------------------------------------------------------
CARTESIAN COORDINATES (ANGSTROEM), LR-TDA-wB97X-D3/def2-TZVP 
MOLECULE: Azulene, STATE: S1
--------------------------------------------------------------------------
18
Azulene S1 LR-TDA-wB97X-D3/def2-TZVP 
  C          -2.22854909622053      1.99584560019162     -0.01311972157042
  C          -1.02749275273831      2.67082577783358     -0.00150865935281
  C          -0.19220118246335     -0.42028779834854     -0.12822593349349
  H           2.65399378178720     -0.21877965949054     -0.12913777526082
  C          -2.48545929461112      0.61827046606560     -0.05485297224904
  C           0.26692717131733      2.08075447304870     -0.03181371655370
  C           2.58532210486969      1.98868070639742     -0.04187643655273
  C           2.07584832304876      0.69491755004590     -0.09328354833271
  C           0.62967740652176      0.74028326727970     -0.08995988581689
  H           1.52961496053804      3.92931115857792      0.03850868042860
  C           1.49248105188215      2.84894590910506     -0.00434373042928
  H           0.31759200283423     -1.37562188331399     -0.17118907582337
  H           3.62645376403206      2.27024826334440     -0.03257454162236
  H          -1.06641698188402      3.75325959009771      0.03557552095904
  C          -1.56960260566854     -0.44169311020572     -0.10659886139875
  H          -3.53132901511115      0.33523078039969     -0.05103786204098
  H          -3.11382320131701      2.62459193750106      0.01497907508075
  H          -2.01805734371719     -1.43049603542961     -0.13602593987107
--------------------------------------------------------------------------
\end{verbatim}
}
\clearpage

\singlespacing
\small
{ 
\begin{verbatim}
--------------------------------------------------------------------------
CARTESIAN COORDINATES (ANGSTROEM), ssUKS-wB97X-D3/def2-TZVP 
MOLECULE: Azulene, STATE: S1
--------------------------------------------------------------------------
18
Azulene S1 ssUKS-wB97X-D3/def2-TZVP 
  C          -2.23331441940160      1.99858376212149     -0.01110652544777
  C          -1.03620511899361      2.66694952553449      0.00664619138974
  C          -0.20119342512852     -0.42137194152240     -0.11858214917013
  H           2.66179790084811     -0.21677389414099     -0.13161386695638
  C          -2.49510768898644      0.61558639867074     -0.06309087294701
  C           0.26968985960959      2.07767590757962     -0.02667553399394
  C           2.59507908123375      1.99136839617786     -0.04813509647471
  C           2.08519336254155      0.69844344418839     -0.09364480666328
  C           0.62970894497863      0.74570927742363     -0.08065562703595
  H           1.54053378733247      3.93241195988801      0.03550167273330
  C           1.50312495141352      2.85153647621169     -0.00666205433339
  H           0.31302453685984     -1.37525906618706     -0.15858932982647
  H           3.63631262181259      2.27273193988856     -0.04520754658557
  H          -1.07341011195830      3.74985620206199      0.04955126050246
  C          -1.57204271726575     -0.44731468934616     -0.11027609693306
  H          -3.54088066491524      0.33298054861560     -0.06724881183459
  H          -3.11823549757124      2.62778475825810      0.01882072393478
  H          -2.01909630930935     -1.43661201232358     -0.14551691425804
--------------------------------------------------------------------------
\end{verbatim}
}
\clearpage

\singlespacing
\small
{ 
\begin{verbatim}
--------------------------------------------------------------------------
CARTESIAN COORDINATES (ANGSTROEM), LR-TDA-wB97X-D3/def2-TZVP 
MOLECULE: Azulene, STATE: T1
--------------------------------------------------------------------------
18
Azulene T1 LR-TDA-wB97X-D3/def2-TZVP 
  C          -2.22511071795127      1.99213045204432     -0.01039339983808
  C          -1.02404942013824      2.67097406691115      0.00312290574962
  C          -0.18899547794710     -0.41955056683409     -0.12072585478253
  H           2.65336528501871     -0.21817099606933     -0.12514451989182
  C          -2.49146831873533      0.61635512609646     -0.05877614148654
  C           0.26916473568309      2.09096368419226     -0.02663926703483
  C           2.58318518170558      1.98898789170058     -0.04617348037469
  C           2.07181351005885      0.69380990051492     -0.08907855469961
  C           0.63667255156661      0.73176552690822     -0.07898312497221
  H           1.53003816326866      3.92914740114695      0.02896933827469
  C           1.48949600629872      2.84834931688863     -0.00909297117681
  H           0.31470541940379     -1.37815071198403     -0.16587172631822
  H           3.62417556589100      2.27045595462616     -0.04286298076370
  H          -1.06981270813167      3.75323179742931      0.04232841463490
  C          -1.56927803638740     -0.43692553830070     -0.10953765329127
  H          -3.53665973280097      0.33398294902109     -0.06001291168791
  H          -3.10826713179759      2.62460957231630      0.01995215039883
  H          -2.01399578190544     -1.42767883350821     -0.14756560663984
--------------------------------------------------------------------------
\end{verbatim}
}
\clearpage 

\singlespacing
\small
{ 
\begin{verbatim}
--------------------------------------------------------------------------
CARTESIAN COORDINATES (ANGSTROEM), UKS-wB97X-D3/def2-TZVP 
MOLECULE: Azulene, STATE: T1
--------------------------------------------------------------------------
18
Azulene T1 UKS-wB97X-D3/def2-TZVP 
  C          -2.23892081150649      1.99200141899195     -0.01205048509043
  C          -1.03426564482758      2.67162432030919      0.00549647010322
  C          -0.20205861135434     -0.41799100432238     -0.11738226410421
  H           2.67009131980379     -0.21808327946592     -0.13039622082502
  C          -2.49858104482062      0.62804813284046     -0.06220887795288
  C           0.26506359972211      2.08834107344936     -0.02695201755057
  C           2.59478994295065      1.97448185328042     -0.04890512165814
  C           2.09565441602603      0.69817611850164     -0.09302341504150
  C           0.63207119527554      0.74678748286110     -0.08023738444291
  H           1.53457500022629      3.92871434031723      0.03367348548676
  C           1.48975671295862      2.84796352520785     -0.00771939531712
  H           0.31300012265987     -1.37201245603509     -0.15669712184129
  H           3.63425577345853      2.26300615658164     -0.04567619758657
  H          -1.07768773177975      3.75395640845884      0.04757285747946
  C          -1.55995928235608     -0.44632417054991     -0.10923916819910
  H          -3.54202248890154      0.33795270723696     -0.06642675814941
  H          -3.12307124128860      2.62299163107219      0.01741433057815
  H          -2.00771213314641     -1.43534726563555     -0.14372809978844
--------------------------------------------------------------------------
\end{verbatim}
}
\clearpage

\singlespacing
\small
{ 
\begin{verbatim}
--------------------------------------------------------------------------
CARTESIAN COORDINATES (ANGSTROEM), RKS-wB97X-D3/def2-TZVP 
MOLECULE: Anthracene, STATE: S0
--------------------------------------------------------------------------
24
Anthracene S0 RKS-wB97X-D3/def2-TZVP 
  C           0.00001320869928      2.46774219628755      1.39890957057547
  C           0.00000254156886      3.63632843602756      0.71195843621507
  C          -0.00001503083516      3.63632684333445     -0.71195473023413
  C          -0.00001021984151      2.46774374689776     -1.39891040881293
  C           0.00000834446883      1.21414699857051     -0.71422116015535
  C           0.00001213476422      1.21414761900378      0.71421872221924
  C           0.00000540389923      0.00000005398311      1.39304669454573
  C           0.00000156169073     -1.21414913473761      0.71421978530851
  C           0.00001529488229     -1.21414969692019     -0.71422117846023
  C           0.00001663825469     -0.00000145508145     -1.39304843424203
  C           0.00000640437915     -2.46774220679168     -1.39891057816741
  C           0.00000091017550     -3.63632821422101     -0.71195686464728
  C          -0.00000702053670     -3.63632638815918      0.71195583480411
  C          -0.00000798440867     -2.46774068084537      1.39890958672137
  H           0.00001840797991     -0.00000215822869     -2.47839074064427
  H          -0.00002151067592     -2.46487108917126      2.48321021377020
  H          -0.00002018488155     -4.58031006498776      1.24339399020687
  H          -0.00000270536241     -4.58031540209017     -1.24339063161729
  H           0.00001019610131     -2.46487411840716     -2.48321232217315
  H           0.00000140297222      0.00000468110833      2.47839044327664
  H          -0.00002335345012      2.46487430086574     -2.48321211127316
  H          -0.00002644040135      4.58031236445434     -1.24339121494521
  H          -0.00000033527744      4.58031355451222      1.24339661684191
  H           0.00002233583463      2.46486981459617      2.48321048088729
--------------------------------------------------------------------------
\end{verbatim}
}
\clearpage

\singlespacing
\small
{ 
\begin{verbatim}
--------------------------------------------------------------------------
CARTESIAN COORDINATES (ANGSTROEM), LR-TDA-wB97X-D3/def2-TZVP 
MOLECULE: Anthracene, STATE: S1
--------------------------------------------------------------------------
24
Anthracene S1 LR-TDA-wB97X-D3/def2-TZVP 
  C          -0.00008544356738      2.46194357509150      1.39114276607323
  C          -0.00005001989480      3.67254174626973      0.69185130317687
  C           0.00008308224183      3.67254092735632     -0.69184912766423
  C           0.00011650714768      2.46194975287613     -1.39114535681880
  C           0.00003786249662      1.23559271359732     -0.71790738374971
  C          -0.00003141741512      1.23559058916296      0.71789527452290
  C          -0.00001823640456     -0.00000564212386      1.39054842001644
  C           0.00002304055573     -1.23559867425824      0.71789884146925
  C          -0.00003973550632     -1.23559040789240     -0.71790465855045
  C          -0.00000519737878      0.00000119905672     -1.39055991689897
  C          -0.00011514502655     -2.46194292939617     -1.39113887760757
  C          -0.00006783035532     -3.67253897773461     -0.69184815644155
  C           0.00008243620800     -3.67254621885875      0.69185286563155
  C           0.00009909013800     -2.46195177496658      1.39114930293387
  H          -0.00001004333137     -0.00000246571522     -2.47641266949548
  H           0.00016982256016     -2.46554722944518      2.47546599324740
  H           0.00016988758332     -4.60708341720121      1.23854223749426
  H          -0.00012443141778     -4.60707443642327     -1.23854136179138
  H          -0.00021075835813     -2.46554286450022     -2.47544956147390
  H          -0.00005549355706      0.00000003140096      2.47640821139303
  H           0.00019225415422      2.46555992958595     -2.47545666595671
  H           0.00012387673846      4.60707844210204     -1.23853813004506
  H          -0.00011944758745      4.60708216637995      1.23853811627594
  H          -0.00016466002338      2.46554396563614      2.47545853425904
--------------------------------------------------------------------------
\end{verbatim}
}
\clearpage

\singlespacing
\small
{ 
\begin{verbatim}
---------------------------------------------------------------------------
CARTESIAN COORDINATES (ANGSTROEM), ssUKS-wB97X-D3/def2-TZVP 
MOLECULE: Anthracene, STATE: S1
---------------------------------------------------------------------------
24
Anthracene S1 ssUKS-wB97X-D3/def2-TZVP 
  C          -0.00002984186672      2.46242004845654      1.38822566395888
  C           0.00003756068208      3.67817315366519      0.68846062612368
  C           0.00002883357902      3.67818959332081     -0.68845583440980
  C           0.00001016169534      2.46244008738358     -1.38823319533463
  C          -0.00003183381978      1.24406481930151     -0.71625877439038
  C          -0.00003970589488      1.24405426473035      0.71624396572339
  C          -0.00004636192562     -0.00000236545987      1.38905716013308
  C          -0.00000524210676     -1.24406098158893      0.71624652954081
  C          -0.00002772947462     -1.24405491713947     -0.71625112936292
  C          -0.00003998896894      0.00000043686537     -1.38906996676218
  C          -0.00001341836578     -2.46242417096454     -1.38822322028487
  C           0.00002071526814     -3.67818180041789     -0.68845467372624
  C           0.00004132062628     -3.67818470040760      0.68846689045494
  C           0.00001998722814     -2.46242458328943      1.38822921383315
  H          -0.00003848551154     -0.00001262581146     -2.47428274906276
  H           0.00001415648792     -2.46568834799510      2.47250043884294
  H           0.00008360002054     -4.61223459951612      1.23628190289834
  H           0.00002767819670     -4.61223143793465     -1.23627110578230
  H          -0.00004569250317     -2.46571784420566     -2.47249949385765
  H          -0.00006520143312      0.00000422806929      2.47427134170408
  H           0.00001432946667      2.46571447123379     -2.47250262045278
  H           0.00007047691190      4.61224505358109     -1.23625378447188
  H           0.00006289734037      4.61222197292342      1.23627647774428
  H          -0.00004821563217      2.46569024519979      2.47249633694081
---------------------------------------------------------------------------
\end{verbatim}
}
\clearpage

\singlespacing
\small
{ 
\begin{verbatim}
---------------------------------------------------------------------------
CARTESIAN COORDINATES (ANGSTROEM), LR-TDA-wB97X-D3/def2-TZVP 
MOLECULE: Anthracene, STATE: T1
---------------------------------------------------------------------------
24
Anthracene T1 LR-TDA-wB97X-D3/def2-TZVP 
  C           0.00007223199261      2.46416527273087      1.38725121238518
  C          -0.00009021663574      3.68000280804332      0.68760733456246
  C          -0.00014988388525      3.68000536194562     -0.68761901145666
  C          -0.00003687914324      2.46415768286926     -1.38724575389064
  C           0.00013457802512      1.24580958391339     -0.71381893661621
  C           0.00018353293146      1.24581064028087      0.71382974404424
  C           0.00020559920293      0.00000113795207      1.39834221706426
  C           0.00013751811537     -1.24580965908436      0.71383286243850
  C           0.00015141327574     -1.24580581330040     -0.71381919029846
  C           0.00022828206653      0.00000218516591     -1.39833735528639
  C           0.00003648306770     -2.46415932153959     -1.38724292736276
  C          -0.00012663423503     -3.68000714752423     -0.68761557473056
  C          -0.00016784786553     -3.68001146545475      0.68761108105325
  C          -0.00002713730863     -2.46416367696351      1.38725082197961
  H           0.00024093129708      0.00000403770853     -2.48274725408071
  H          -0.00005140115637     -2.46665004362093      2.47151843733225
  H          -0.00026575674647     -4.61358510714189      1.23629952140915
  H          -0.00021412504214     -4.61356937293056     -1.23632606747183
  H           0.00005755143372     -2.46664331114009     -2.47151247564977
  H           0.00019438216442      0.00000573441712      2.48274997705852
  H          -0.00008740156299      2.46663177326516     -2.47151408718804
  H          -0.00032012221197      4.61357440710571     -1.23631680655434
  H          -0.00019567646284      4.61357432463533      1.23630282380254
  H           0.00009057868354      2.46665996866713      2.47151940745636
---------------------------------------------------------------------------
\end{verbatim}
}
\clearpage

\singlespacing
\small
{ 
\begin{verbatim}
--------------------------------------------------------------------------
CARTESIAN COORDINATES (ANGSTROEM), UKS-wB97X-D3/def2-TZVP 
MOLECULE: Anthracene, STATE: T1
--------------------------------------------------------------------------
24
Anthracene T1 UKS-wB97X-D3/def2-TZVP 
  C           0.00004319450625      2.46474621803113      1.38585248086369
  C          -0.00002175436833      3.68320152353206      0.68580911714451
  C          -0.00008410525157      3.68321551847266     -0.68581359283358
  C          -0.00004366189108      2.46476186772384     -1.38586497598170
  C           0.00003964052089      1.25351207557391     -0.71484883867648
  C           0.00006459464470      1.25350456653223      0.71483140186611
  C           0.00007822001227     -0.00000055183152      1.40332623226531
  C           0.00004379916117     -1.25351187374532      0.71484926734339
  C           0.00007572457515     -1.25351124239474     -0.71483350506354
  C           0.00010940646887     -0.00000393465121     -1.40331722231138
  C           0.00005432625311     -2.46475430216562     -1.38586278244843
  C          -0.00004485497120     -3.68320640855431     -0.68581186686869
  C          -0.00007613679975     -3.68320985055378      0.68581114970091
  C          -0.00002684249077     -2.46475246053543      1.38586572244044
  H           0.00010927501233     -0.00001248116393     -2.48740722561815
  H          -0.00005936088277     -2.46806172460481      2.47022782990214
  H          -0.00013554475489     -4.61674395894992      1.23483307448853
  H          -0.00009337474153     -4.61674358443474     -1.23483629383017
  H           0.00008646669407     -2.46804891436435     -2.47022690873576
  H           0.00008927809391      0.00001961560429      2.48741929431704
  H          -0.00006589550538      2.46806323402155     -2.47023012887527
  H          -0.00016814228418      4.61675778507624     -1.23482381312030
  H          -0.00003829530425      4.61673477138368      1.23483605071845
  H           0.00006404330295      2.46804411199808      2.47021553331295
---------------------------------------------------------------------------
\end{verbatim}
}
\clearpage 

\singlespacing
\small
{ 
\begin{verbatim}
--------------------------------------------------------------------------
CARTESIAN COORDINATES (ANGSTROEM), RKS-wB97X-D3/def2-TZVP 
MOLECULE: Fluoranthene, STATE: S0
--------------------------------------------------------------------------
26
Fluoranthene S0 RKS-wB97X-D3/def2-TZVP 
  C          -0.00000011226816      2.41406411625359      2.15290584494619
  C           0.00000463642165      2.38064755526854      0.73431618755689
  C          -0.00000089614562      1.16605281694668      0.10915862698117
  C          -0.00000704197307      0.00000457080402      0.90308869287669
  C          -0.00000488880609     -0.00000107960683      2.29210331864859
  C          -0.00000862747295      1.27446062706540      2.91635957694591
  H          -0.00001733374933      1.34652678415203      3.99814165260960
  C           0.00000163313867     -1.27446327859122      2.91635456330409
  C           0.00000821149695     -2.41406557696013      2.15289782542634
  C           0.00000984607386     -2.38063998873734      0.73430992285280
  C          -0.00000212445984     -1.16604095184486      0.10915855635761
  C          -0.00000780574072     -0.70623211988728     -1.29385237929598
  C          -0.00000272526975      0.70624040465118     -1.29385282338457
  C           0.00000455087086      1.40534219682415     -2.48564524925240
  C           0.00000322710440      0.69419090845338     -3.68093090830289
  C          -0.00000630164975     -0.69420091231229     -3.68092614897398
  C          -0.00001224603585     -1.40534647406307     -2.48563658463974
  H          -0.00001228320293     -2.48916389977311     -2.49411360993036
  H          -0.00001055566684     -1.22937530235895     -4.62281347647520
  H           0.00000846399431      1.22936248880579     -4.62282037104224
  H           0.00001645567734      2.48915909640384     -2.49413113858268
  H           0.00001355467590     -3.31109150996344      0.17831972869194
  H           0.00001300387547     -3.37837681451427      2.64707493529495
  H           0.00000070900088     -1.34653377953341      3.99813694645653
  H           0.00001059307506      3.31110666071588      0.17833691737888
  H          -0.00000194296446      3.37837346180173      2.64708539355187
---------------------------------------------------------------------------
\end{verbatim}
}
\clearpage 

\singlespacing
\small
{ 
\begin{verbatim}
--------------------------------------------------------------------------
CARTESIAN COORDINATES (ANGSTROEM), LR-TDA-wB97X-D3/def2-TZVP 
MOLECULE: Fluoranthene, STATE: S1
--------------------------------------------------------------------------
26
Fluoranthene S1 LR-TDA-wB97X-D3/def2-TZVP 
  C          -0.00000966011503      2.43461651459854      2.11706397209533
  C          -0.00019369665697      2.43361213473839      0.73175921532997
  C          -0.00019121711821      1.18705371102140      0.07891082897972
  C          -0.00014345256329      0.00001035426494      0.88537754468138
  C           0.00013092664828     -0.00000461189513      2.30095679216687
  C           0.00019368996396      1.27105754259584      2.89998329821790
  H           0.00041151375889      1.35617587569124      3.98003807082562
  C           0.00028809949669     -1.27109176663169      2.89994470015201
  C           0.00007728484929     -2.43463192616309      2.11699673524806
  C          -0.00023080077117     -2.43359307755456      0.73169291877749
  C          -0.00026205750927     -1.18701074624997      0.07889209896961
  C          -0.00022120894561     -0.73707535931945     -1.24787459598498
  C          -0.00017293107718      0.73713379471349     -1.24787169800766
  C           0.00001704168513      1.42939495023243     -2.48071613382684
  C           0.00013738932612      0.71612976232988     -3.64019724184581
  C           0.00003857164923     -0.71614076692366     -3.64017852799595
  C          -0.00014111656780     -1.42937243894971     -2.48068508618401
  H          -0.00019192308841     -2.51228358529152     -2.49531523175434
  H           0.00011529289700     -1.23338001942982     -4.59184619050567
  H           0.00030530562795      1.23332692760679     -4.59188533250620
  H           0.00010939822898      2.51230342341948     -2.49538442151057
  H          -0.00041051171450     -3.36432665969719      0.17985165375727
  H           0.00017714479630     -3.38889080243700      2.63173903909648
  H           0.00053501406974     -1.35623880714450      3.97999787330540
  H          -0.00034615246193      3.36436439412068      0.17994725721233
  H          -0.00002194440820      3.38886118235419      2.63182846130660
---------------------------------------------------------------------------
\end{verbatim}
}
\clearpage

\singlespacing
\small
{ 
\begin{verbatim}
--------------------------------------------------------------------------
CARTESIAN COORDINATES (ANGSTROEM), ssUKS-wB97X-D3/def2-TZVP 
MOLECULE: Fluoranthene, STATE: S1
--------------------------------------------------------------------------
26
Fluoranthene S1 ssUKS-wB97X-D3/def2-TZVP 
  C           0.00039254814164      2.44184319829072      2.11692829554047
  C           0.00018901989714      2.43691924984179      0.73702505208737
  C          -0.00005363541134      1.18541349948752      0.07958369679072
  C          -0.00003084058270     -0.00000440685558      0.88015756956352
  C          -0.00001882252178      0.00000110029846      2.29564075269242
  C           0.00024168023542      1.27048752506288      2.89762132125987
  H           0.00034729799709      1.35320237301659      3.97799734493910
  C          -0.00024755036781     -1.27048142559181      2.89765629511052
  C          -0.00038547831756     -2.44181741061934      2.11696134060791
  C          -0.00021019773466     -2.43693011585433      0.73704189095972
  C          -0.00000421558093     -1.18544899447033      0.07954744587602
  C           0.00010001387897     -0.74183426086552     -1.24640232038709
  C          -0.00014496555168      0.74179821290500     -1.24638558948389
  C          -0.00052488511668      1.43361821994367     -2.48246368970213
  C          -0.00032577190280      0.71990003050507     -3.63824477293568
  C           0.00036141098620     -0.71988091224491     -3.63826856220542
  C           0.00051266875488     -1.43361009509479     -2.48250168588471
  H           0.00090513906338     -2.51659682122192     -2.49873256882279
  H           0.00068953079750     -1.23478291498132     -4.59130753806116
  H          -0.00062277076676      1.23481463024886     -4.59128070043977
  H          -0.00093392510852      2.51660618850884     -2.49866290405678
  H          -0.00028937857454     -3.36699832718745      0.18305959531746
  H          -0.00059198999194     -3.39485851138611      2.63352240305045
  H          -0.00030132038468     -1.35320525736022      3.97802736943863
  H           0.00027996939955      3.36696837690901      0.18301209698676
  H           0.00066646876262      3.39487684871521      2.63349386175849
---------------------------------------------------------------------------
\end{verbatim}
}
\clearpage

\singlespacing
\small
{ 
\begin{verbatim}
--------------------------------------------------------------------------
CARTESIAN COORDINATES (ANGSTROEM), LR-TDA-wB97X-D3/def2-TZVP 
MOLECULE: Fluoranthene, STATE: T1
--------------------------------------------------------------------------
26
Fluoranthene T1 LR-TDA-wB97X-D3/def2-TZVP 
  C           0.00006837521674      2.45420659023609      2.12105036824349
  C          -0.00006091429711      2.43147178598168      0.75518787697502
  C          -0.00007530114045      1.15855695201987      0.10030165502676
  C          -0.00006329379978     -0.00003600527374      0.89148447625254
  C          -0.00001693544577     -0.00002631894022      2.29595010138044
  C           0.00011977136673      1.26396287644761      2.91392578913772
  H           0.00018386446037      1.34463512885613      3.99416891030116
  C          -0.00004220808962     -1.26397609271865      2.91395833743102
  C          -0.00007018525329     -2.45420981273343      2.12110705516322
  C          -0.00003265999645     -2.43147882011748      0.75522673754288
  C          -0.00001350597023     -1.15860047523751      0.10032695882642
  C           0.00003217279166     -0.71348774939240     -1.28267060106975
  C          -0.00004399642946      0.71346607469277     -1.28265783954267
  C          -0.00007225783213      1.40761729371914     -2.47765725929322
  C          -0.00003371293967      0.69381903831245     -3.67632524588612
  C           0.00006325608896     -0.69376708275205     -3.67633345308154
  C           0.00011800393081     -1.40759504189534     -2.47767430596696
  H           0.00020435165124     -2.49163936059022     -2.49046558859042
  H           0.00011512643288     -1.23018969665899     -4.61720846599530
  H          -0.00006518792199      1.23024492873274     -4.61720144734662
  H          -0.00013960126471      2.49166433651270     -2.49041221828498
  H          -0.00002162825359     -3.35339560850890      0.18753892173309
  H          -0.00010684755781     -3.40812675459451      2.63488965127650
  H          -0.00002838144460     -1.34461316335450      3.99420429661046
  H          -0.00012071936571      3.35337436523466      0.18747310243300
  H           0.00010241506296      3.40812261202212      2.63483818672386
---------------------------------------------------------------------------
\end{verbatim}
}
\clearpage

\singlespacing
\small
{ 
\begin{verbatim}
--------------------------------------------------------------------------
CARTESIAN COORDINATES (ANGSTROEM), UKS-wB97X-D3/def2-TZVP 
MOLECULE: Fluoranthene, STATE: T1
--------------------------------------------------------------------------
26
Fluoranthene T1 UKS-wB97X-D3/def2-TZVP 
  C           2.54063830898191      1.99278674291527      0.00001800249276
  C           2.47290874547250      0.64269020816034      0.00001680904942
  C           1.15427465193029      0.02342492801102      0.00002092954680
  C          -0.00289816097118      0.89318930430772      0.00004898895251
  C           0.06077734636168      2.29028649591384      0.00001716820230
  C           1.37478365715393      2.84650065758477      0.00001800313436
  H           1.51043680760115      3.92091055936307      0.00002813863558
  C          -1.16261454740318      2.96671188536367     -0.00003592502629
  C          -2.36509857988537      2.24048940788625     -0.00002431474675
  C          -2.39881721082576      0.85836415301865      0.00001619750273
  C          -1.17552952502776      0.16252723277907      0.00005995753307
  C          -0.77050668516989     -1.24551172908510      0.00003285376692
  C           0.67016996367698     -1.29349213033127     -0.00000607888436
  C           1.32865420036433     -2.53724775425953     -0.00004688104145
  C           0.57838016703810     -3.69302590974563     -0.00004512588682
  C          -0.82116265549094     -3.63571512680813      0.00000475189601
  C          -1.49738610296962     -2.41673097551088      0.00004211854760
  H          -2.58111971796006     -2.39784228867596      0.00008282526920
  H          -1.38905885529358     -4.55836269411557      0.00001186471180
  H           1.07139662581730     -4.65722453817775     -0.00006957806505
  H           2.41147716865713     -2.58462362912065     -0.00007075137715
  H          -3.34777102978240      0.33592065424020      0.00000152070071
  H          -3.30004943949433      2.78840867307228     -0.00006826904412
  H          -1.18837717385638      4.05011620378244     -0.00009808746948
  H           3.37121231818407      0.03899106784612      0.00002720210490
  H           3.51154772289107      2.47440960158574      0.00001767949480
---------------------------------------------------------------------------
\end{verbatim}
}
\clearpage

\singlespacing
\small
{ 
\begin{verbatim}
--------------------------------------------------------------------------
CARTESIAN COORDINATES (ANGSTROEM), RKS-wB97X-D3/def2-TZVP 
MOLECULE: Pyruvic Acid, STATE: S0
--------------------------------------------------------------------------
10
Pyruvic Acid S0 RKS-wB97X-D3/def2-TZVP 
  C          -6.25743668705252      0.55956230941167     -0.01997086341407
  C          -4.99036299319624      1.34063890614173      0.02868059553956
  H          -7.10587522169087      1.22988340885836     -0.12936316423591
  H          -6.21449818089323     -0.14751360720306     -0.85056697399784
  H          -6.34991579802749     -0.04036126804752      0.88695841814805
  H          -2.92087852366914      2.23115889897213      0.13569508876997
  C          -3.67831156046844      0.53773179929782      0.20557024384831
  O          -3.65362291899794     -0.65267218865344      0.30268194336662
  O          -4.90425955982614      2.54021011572648     -0.05861763464816
  O          -2.60668796357794      1.31621472919584      0.23573919082347
---------------------------------------------------------------------------
\end{verbatim}
}
\clearpage 

\singlespacing
\small
{ 
\begin{verbatim}
--------------------------------------------------------------------------
CARTESIAN COORDINATES (ANGSTROEM), LR-TDA-wB97X-D3/def2-TZVP 
MOLECULE: Pyruvic Acid, STATE: S1
--------------------------------------------------------------------------
10
Pyruvic Acid S1 LR-TDA-wB97X-D3/def2-TZVP 
  C          -6.25595497542124      0.52687250676630     -0.07703536227544
  C          -4.95189491231194      1.26811073412121     -0.09382212742413
  H          -6.98889128669253      1.06186650049514      0.53042391999658
  H          -6.65158461358880      0.43345910416888     -1.09043666192035
  H          -6.05797003526372     -0.45812621244801      0.34646046674758
  H          -2.85082995582522      2.20698748250151     -0.01073915157423
  C          -3.73101168749584      0.60926420747789      0.42327379079608
  O          -3.72343356813114     -0.52230477646972      0.86241034265944
  O          -4.85157437524769      2.44093605235987     -0.52634199659298
  O          -2.61870399742182      1.34778750472692      0.37261362378743
---------------------------------------------------------------------------
\end{verbatim}
}
\clearpage

\singlespacing
\small
{ 
\begin{verbatim}
--------------------------------------------------------------------------
CARTESIAN COORDINATES (ANGSTROEM), ssUKS-wB97X-D3/def2-TZVP 
MOLECULE: Pyruvic Acid, STATE: S1
--------------------------------------------------------------------------
10
Pyruvic Acid S1 ssUKS-wB97X-D3/def2-TZVP 
  C          -6.27357365203086      0.53599663794576      0.03681442342887
  C          -4.96982766989006      1.26187754040495      0.16138159811783
  H          -6.88247884751414      0.97226062897062     -0.75759521970126
  H          -6.03161162972047     -0.49937684854263     -0.20018095841927
  H          -6.83418312815361      0.58716630969804      0.97234772196002
  H          -2.82221712641473      2.23860146236311      0.30822644462249
  C          -3.68803721406437      0.57346494099341     -0.03121023502220
  O          -3.61945510070093     -0.60053944196694     -0.29243660851016
  O          -4.97800794775534      2.51713897846565      0.43946451351094
  O          -2.58245709115543      1.32826289536802      0.09999516421276
---------------------------------------------------------------------------
\end{verbatim}
}
\clearpage

\singlespacing
\small
{ 
\begin{verbatim}
--------------------------------------------------------------------------
CARTESIAN COORDINATES (ANGSTROEM), LR-TDA-wB97X-D3/def2-TZVP 
MOLECULE: Pyruvic Acid, STATE: T1
--------------------------------------------------------------------------
10
Pyruvic Acid T1 LR-TDA-wB97X-D3/def2-TZVP 
  C          -6.25815776539793      0.52260786212547     -0.07967309668295
  C          -4.95425604039978      1.26018435277493     -0.10384226055245
  H          -6.97848597745011      1.04487110544684      0.55356888131821
  H          -6.67277011726146      0.44895223250020     -1.08643141898497
  H          -6.06250403070566     -0.47107448519210      0.32116621654018
  H          -2.82898953132771      2.21054185583364     -0.01557778062460
  C          -3.73174029105287      0.61949737793844      0.43126092730593
  O          -3.73207583216391     -0.50573872481830      0.88086695143940
  O          -4.84784213665401      2.43050753466707     -0.54434458870581
  O          -2.61502768498649      1.35450399242382      0.37981301314706
---------------------------------------------------------------------------
\end{verbatim}
}

\clearpage 

\singlespacing
\small
{ 
\begin{verbatim}
--------------------------------------------------------------------------
CARTESIAN COORDINATES (ANGSTROEM), UKS-wB97X-D3/def2-TZVP 
MOLECULE: Pyruvic Acid, STATE: T1
--------------------------------------------------------------------------
10
Pyruvic Acid T1 UKS-wB97X-D3/def2-TZVP 
  C          -6.27297062906535      0.53394561040282      0.03918077996314
  C          -4.97380220259966      1.26904056456308      0.16264528252139
  H          -6.88522792450075      0.96573748414406     -0.75471354885083
  H          -6.02548110700346     -0.50005076324886     -0.19833658178522
  H          -6.83238682734950      0.58074050188149      0.97532761834201
  H          -2.82836972566947      2.23909383288121      0.30679878024514
  C          -3.69048119648865      0.57853872031910     -0.03322502240940
  O          -3.62687509786077     -0.59609147813147     -0.29606509587536
  O          -4.96156718506635      2.51415334956445      0.43877688689987
  O          -2.58468751179600      1.32974528132411      0.09641774514926
---------------------------------------------------------------------------
\end{verbatim}
}

\clearpage

\singlespacing
\small
{ 
\begin{verbatim}
--------------------------------------------------------------------------
CARTESIAN COORDINATES (ANGSTROEM), RKS-wB97X-D3/def2-TZVP 
MOLECULE: Rubrene, STATE: S0
--------------------------------------------------------------------------
70
Rubrene S0 RKS-wB97X-D3/def2-TZVP 
  C          -0.67851983227284     -4.86499065465507     -0.21899785341218
  C           0.67838171595685     -4.86502205226822      0.21873611657937
  C           1.33721751928414     -3.69980872665932      0.40083644092691
  C           0.69846122167679     -2.43289639311881      0.16275278687546
  C          -0.69857876791195     -2.43285921298358     -0.16291605365173
  C          -1.33736110492810     -3.69976289619355     -0.40098762508465
  C           1.39905903202752     -1.23377877413189      0.20986274423248
  C           0.72043904595538      0.00000634314109     -0.00010549850773
  C          -0.72045053270941      0.00004348777170     -0.00012145370177
  C          -1.39914258886431     -1.23371644823179     -0.21001522572109
  C           1.39913344498409      1.23376228752660     -0.21003266897474
  C           0.69856934075479      2.43290226471234     -0.16295624683011
  C          -0.69847490290436      2.43294567236876      0.16270348565271
  C          -1.39908145449975      1.23383229059576      0.20983799136651
  C           1.33738103853942      3.69979841077902     -0.40100363926151
  C           0.67856432638805      4.86504168911355     -0.21901863096221
  C          -0.67835755643838      4.86509201013705      0.21866452984985
  C          -1.33721420530041      3.69988064986861      0.40072937903876
  C          -2.82753032124274     -1.28133609847823     -0.64150313846605
  C          -2.82741477410024      1.28138953122196      0.64147955064362
  C           2.82739712806208     -1.28136314971886      0.64146928107164
  C          -3.13781598250110     -0.94644802630898     -1.95538711639940
  C          -4.44112422179301     -1.03294032715999     -2.41985318017669
  C          -5.45302908740788     -1.45472670519979     -1.57237060470908
  C          -5.15115124772804     -1.80063427598546     -0.26307489722907
  C          -3.84684468431916     -1.72402097328954      0.19472678998115
  C          -3.84686897703301      1.72407237384826     -0.19457695065656
  C          -5.15113381292649      1.80049743475222      0.26337460697221
  C          -5.45281514477923      1.45447053201231      1.57268366967356
  C          -4.44077088220186      1.03271245978603      2.42001257224639
  C          -3.13751971216745      0.94635025883458      1.95536725497323
  C           3.13750635762085     -0.94655400437269      1.95541444856821
  C           4.44077354420908     -1.03295673235980      2.42001989367691
  C           5.45281358280767     -1.45453476862132      1.57259774092261
  C           5.15111852887537     -1.80035218516374      0.26323878630652
  C           3.84684656057318     -1.72388050421484     -0.19468127225845
  C           2.82752480407910      1.28134855786514     -0.64149652389739
  C           3.84685738080392      1.72399778599775      0.19472723914048
  C           5.15116922051023      1.80053993697792     -0.26307697029114
  C           5.45301862970723      1.45467063319346     -1.57238775000837
  C           4.44109516988805      1.03293345122141     -2.41987501002397
  C           3.13779233584291      0.94647123672211     -1.95539093520310
  H          -1.18398971197521     -5.80673266766609     -0.39565018627056
  H           1.18383384935844     -5.80678484705807      0.39532906084878
  H           2.36966369313072     -3.70897506283325      0.72102547306224
  H          -2.36981325253557     -3.70893250477614     -0.72115566062194
  H           2.36984487646605      3.70892841680616     -0.72114692264674
  H           1.18406372663617      5.80677690536804     -0.39562526351735
  H          -1.18381005343891      5.80686256548660      0.39522072447233
  H          -2.36967641290654      3.70905984239791      0.72087456797123
  H          -2.34567877529490     -0.61433616636449     -2.61689495952038
  H          -4.66549199088952     -0.76964434182365     -3.44667549182220
  H          -6.47376944028665     -1.51546926931849     -1.93002796208001
  H          -5.93605854833118     -2.12757790040276      0.40813936440003
  H          -3.61365963414952     -1.99230782676553      1.21805170336676
  H          -3.61383886948579      1.99246806655330     -1.21790785409011
  H          -5.93615731513629      2.12740593798730     -0.40772024988169
  H          -6.47351440230765      1.51509128087525      1.93047861270229
  H          -4.66499243744821      0.76932737356383      3.44684394483088
  H          -2.34527685988522      0.61422719261020      2.61674465494741
  H           2.34526801568054     -0.61456253227546      2.61686276303758
  H           4.66500472513974     -0.76975636429048      3.44689815857644
  H           6.47352477265936     -1.51517740248238      1.93035526410285
  H           5.93613634440149     -2.12712284168272     -0.40793179234928
  H           3.61381651465349     -1.99208506616612     -1.21806197986636
  H           3.61369843650706      1.99226985507970      1.21806203848011
  H           5.93610117380216      2.12741869880646      0.40813911142213
  H           6.47375780696237      1.51538872394494     -1.93005070263602
  H           4.66544384833390      0.76965322064839     -3.44670335183493
  H           2.34563978582267      0.61437032444538     -2.61688912835509
---------------------------------------------------------------------------
\end{verbatim}
}

\clearpage

\singlespacing
\small
{ 
\begin{verbatim}
--------------------------------------------------------------------------
CARTESIAN COORDINATES (ANGSTROEM), ssUKS-wB97X-D3/def2-TZVP 
MOLECULE: Rubrene, STATE: S1
--------------------------------------------------------------------------
70
Rubrene S1 ssUKS-wB97X-D3/def2-TZVP 
  C          -0.65673980403561     -4.88754851146326     -0.22814217572445
  C           0.65654163401815     -4.88750511991251      0.22970298467136
  C           1.31529732680900     -3.69053458456717      0.43161812547268
  C           0.68702634060261     -2.46001479365044      0.18820679276354
  C          -0.68715202014477     -2.46004940075700     -0.18736111631694
  C          -1.31545949191211     -3.69061580979575     -0.43044459546862
  C           1.38808135821901     -1.21222772396490      0.26970513526142
  C           0.72935211730785     -0.00005368066397      0.00022466929110
  C          -0.72948642742426     -0.00006996537474      0.00027520734341
  C          -1.38815471094428     -1.21226640141535     -0.26925020297420
  C           1.38798510214396      1.21210936522095     -0.26947427494577
  C           0.68688517133329      2.45986382982135     -0.18797930004259
  C          -0.68725343284209      2.45990027992299      0.18775601551664
  C          -1.38822685161654      1.21210836575779      0.26969635729091
  C           1.31507972071392      3.69039181886553     -0.43161257233532
  C           0.65632989769403      4.88736357868313     -0.22966320918048
  C          -0.65685018285179      4.88741369539652      0.22847478698075
  C          -1.31550369564973      3.69047217848829      0.43094566259069
  C          -2.77820159047046     -1.20640783383596     -0.79558516433628
  C          -2.77839505763519      1.20634848536025      0.79573643315015
  C           2.77837038498029     -1.20656027554338      0.79543676469527
  C          -3.01408026160049     -0.64002751246172     -2.04614334696478
  C          -4.28756215545665     -0.62851275672870     -2.59198030420311
  C          -5.34864266079256     -1.17836084727535     -1.89041484092499
  C          -5.12496400385720     -1.74765058223354     -0.64501659543178
  C          -3.84962669503807     -1.77009676378697     -0.10674981567281
  C          -3.84951997137062      1.77035797052881      0.10670185957517
  C          -5.12499517350570      1.74816488374534      0.64465359801987
  C          -5.34913622238624      1.17872515301606      1.88990053424940
  C          -4.28835925321326      0.62852114600416      2.59165317026830
  C          -3.01473631576001      0.63983114687446      2.04614673176095
  C           3.01495065220189     -0.64054770640185      2.04602828267431
  C           4.28871238004854     -0.62934534070967      2.59123329884864
  C           5.34934676098657     -1.17927318855382      1.88905099477000
  C           5.12492928481067     -1.74828932164932      0.64366065671367
  C           3.84934647218522     -1.77033397861405      0.10597745565719
  C           2.77819150546000      1.20647003093430     -0.79540253598368
  C           3.84920976335826      1.77017867562353     -0.10596394257176
  C           5.12476085352352      1.74819262111939     -0.64372420094767
  C           5.34905820722898      1.17931428139101     -1.88919687149546
  C           4.28837642274804      0.62949068510722     -2.59138170596735
  C           3.01465817235131      0.64058395032580     -2.04606534787307
  H          -1.17068016633291     -5.82244255144538     -0.41514996304069
  H           1.17045304172685     -5.82235442444420      0.41697807621898
  H           2.34134492194113     -3.70137156900142      0.77243990027975
  H          -2.34148365067926     -3.70156505918389     -0.77134807254110
  H           2.34106134874899      3.70122892493900     -0.77265297181916
  H           1.17018130803098      5.82221341368874     -0.41712652109660
  H          -1.17078183553154      5.82229222374083      0.41557296444861
  H          -2.34145730205578      3.70139503697554      0.77205055706928
  H          -2.18515333519235     -0.20044040609096     -2.58889312637532
  H          -4.44968316492943     -0.18483571226188     -3.56718169647068
  H          -6.34686639265566     -1.16323973939026     -2.31094013794243
  H          -5.95024661333648     -2.17171928554348     -0.08578644750478
  H          -3.68349672450615     -2.20334669052625      0.87229902469338
  H          -3.68302968181025      2.20367822371425     -0.87225381843559
  H          -5.95002860686558      2.17251192082844      0.08526768262836
  H          -6.34747557816914      1.16375277616313      2.31015932362113
  H          -4.45083929538391      0.18468280021484      3.56672005210303
  H          -2.18604358199436      0.20000416610783      2.58905745299275
  H           2.18635601169919     -0.20097733271182      2.58930948137505
  H           4.45138400729523     -0.18585627738175      3.56642373526900
  H           6.34778487767570     -1.16441314718524      2.30908131016003
  H           5.94984584172747     -2.17246293182308      0.08396366261245
  H           3.68265270810905     -2.20336346439395     -0.87307374877745
  H           3.68253376885474      2.20310756318162      0.87313761243162
  H           5.94973328849834      2.17227792451165     -0.08404519629220
  H           6.34746405825561      1.16450289576715     -2.30930834252272
  H           4.45099930158804      0.18614019701702     -3.56664466718553
  H           2.18601789507398      0.20105048170609     -2.58929952410352
---------------------------------------------------------------------------
\end{verbatim}
}

\clearpage

\singlespacing
\small
{ 
\begin{verbatim}
--------------------------------------------------------------------------
CARTESIAN COORDINATES (ANGSTROEM), UKS-wB97X-D3/def2-TZVP 
MOLECULE: Rubrene, STATE: T1
--------------------------------------------------------------------------
70
Rubrene T1 UKS-wB97X-D3/def2-TZVP 
  C          -0.66223824011280     -4.89725309908152     -0.20785657989654
  C           0.66211513416364     -4.89752656679080      0.20089691077080
  C           1.32582251175316     -3.69726688243036      0.38377501075814
  C           0.68673572761968     -2.47234444723398      0.16940024795813
  C          -0.68692981355412     -2.47212998862534     -0.17306733240587
  C          -1.32601424202716     -3.69676102778956     -0.38903781415897
  C           1.39920016588128     -1.21366668311544      0.24685786818226
  C           0.73503143857534      0.00002470561171     -0.00081511447386
  C          -0.73499281682388      0.00010127186159     -0.00111545758878
  C          -1.39936737452856     -1.21329157240923     -0.24879138935427
  C           1.39951881997338      1.21348989034916     -0.24814825463781
  C           0.68703404749726      2.47234357756558     -0.17226323515082
  C          -0.68680110792014      2.47251617112516      0.16943655961898
  C          -1.39925211741307      1.21380819790592      0.24649196707920
  C           1.32626297334190      3.69701896522373     -0.38753944527707
  C           0.66237586384109      4.89749348969886     -0.20652817713783
  C          -0.66226484902202      4.89768716527242      0.20128047533512
  C          -1.32611043395357      3.69738356319032      0.38349862531608
  C          -2.80172042564894     -1.23468954882856     -0.74244086361817
  C          -2.80085683115521      1.23510175198850      0.74221562813667
  C           2.80080877234920     -1.23499697010892      0.74257688346166
  C          -3.06398380873354     -0.75898431496501     -2.02423399006231
  C          -4.34907304464021     -0.78729691478406     -2.54271481987584
  C          -5.39344278453676     -1.28880437660997     -1.78262073964670
  C          -5.14210030497054     -1.77179553146597     -0.50636756471324
  C          -3.85573702989204     -1.75364322595303      0.00506585284969
  C          -3.85652595643113      1.75303213118928     -0.00364259043716
  C          -5.14196740548621      1.77059538490676      0.51012800427049
  C          -5.39073757395723      1.28812532293562      1.78708300537170
  C          -4.34468672226649      0.78760581472686      2.54551529102017
  C          -3.06053768454508      0.75981640673822      2.02468974950813
  C           3.06069973135333     -0.75935303900644      2.02488452729465
  C           4.34491385716363     -0.78706537813910      2.54554600868884
  C           5.39084196140265     -1.28787418136108      1.78712686330412
  C           5.14188526431870     -1.77077249505238      0.51036950861607
  C           3.85635614433868     -1.75330173731883     -0.00321715433603
  C           2.80179654570065      1.23475301074667     -0.74200150883306
  C           3.85605928306939      1.75381725431171      0.00508378060374
  C           5.14230248602085      1.77157880846943     -0.50670230882908
  C           5.39329880300296      1.28812110497736     -1.78285250249288
  C           4.34867253710044      0.78646455691720     -2.54250159650694
  C           3.06371759297602      0.75852004432271     -2.02366279023908
  H          -1.18459256769823     -5.83119816581403     -0.37575936591629
  H           1.18449327771240     -5.83168614818386      0.36750734916323
  H           2.36271548719527     -3.70732369260364      0.69037381213836
  H          -2.36291764690429     -3.70650907035870     -0.69562490084780
  H           2.36335712676827      3.70678546209775     -0.69348983788867
  H           1.18485712866341      5.83146847249024     -0.37386069772594
  H          -1.18474544236338      5.83183609017713      0.36767286706466
  H          -2.36320706532297      3.70735225372341      0.68941882698636
  H          -2.24765031101274     -0.35898045843628     -2.61471121046362
  H          -4.53380930858271     -0.41296763113129     -3.54264909992730
  H          -6.40027800576741     -1.30417926626258     -2.18211620042429
  H          -5.95377369214255     -2.15796455788976      0.09834735997858
  H          -3.66582695203299     -2.12247762529125      1.00595711662233
  H          -3.66868738608521      2.12137039272616     -1.00510502806115
  H          -5.95498700063032      2.15581780934560     -0.09338529898383
  H          -6.39686742650767      1.30309923315201      2.18837033488766
  H          -4.52734106298314      0.41370798249769      3.54598876323108
  H          -2.24288693957920      0.36058325596777      2.61387267783945
  H           2.24315942175119     -0.35986863187140      2.61405806311899
  H           4.52773677515154     -0.41285952405193      3.54587923907902
  H           6.39703160283821     -1.30270154396271      2.18827025281190
  H           5.95478646323178     -2.15626788503943     -0.09312781643214
  H           3.66837404743675     -2.12197673284871     -1.00452797553074
  H           3.66647462685229      2.12306585895656      1.00588907584161
  H           5.95417947166966      2.15783719538843      0.09768280722205
  H           6.40003338074568      1.30326010471388     -2.18261437215497
  H           4.53307933975186      0.41177590169230     -3.54236557540291
  H           2.24717356401998      0.35838031185140     -2.61376270469799
---------------------------------------------------------------------------
\end{verbatim}
}

\clearpage
